\documentclass[final]{IEEEtran}

\usepackage[utf8]{inputenc}
\usepackage[T1]{fontenc}

\usepackage{times}
\usepackage{epsfig}
\usepackage{graphicx}
\usepackage{amsmath}
\usepackage{amssymb}

\usepackage{multirow}
\usepackage{color}
\usepackage{graphicx}
\graphicspath{{figures/}}

\usepackage[noadjust]{cite}

\usepackage{amssymb,amsfonts}
\usepackage{amsmath,amssymb}
\usepackage{amsthm}
\usepackage{float}
\usepackage{amsmath,amsfonts}
\usepackage{comment}
\usepackage{verbatim}

\usepackage[all,dvips,arc,curve,color,frame]{xy}
\usepackage{algorithm}
\usepackage{algorithmicx}
\usepackage{bbm}
\usepackage{wrapfig}
\usepackage{dsfont}
\usepackage{subcaption}
\usepackage{mathtools}

\usepackage{algpseudocode}

\usepackage[pagebackref=true,breaklinks=true,letterpaper=true,colorlinks,bookmarks=false]{hyperref}

\newcommand{\argmin}{\operatornamewithlimits{argmin}}

\newcommand{\reid}{re-ID}

\newtheorem{thm}{Theorem}[section]

\newtheorem*{definition*}{Definition}

\ifCLASSINFOpdf
\else
\fi

\makeatletter
\def\ps@IEEEtitlepagestyle{%
  \def\@oddfoot{\mycopyrightnotice}%
  \def\@evenfoot{}%
}
\def\mycopyrightnotice{%
  {\footnotesize 1520-9210 (c) 2018 IEEE. Citation information: DOI 10.1109/TMM.2018.2865860, IEEE Transactions on Multimedia.\hfill}
  \gdef\mycopyrightnotice{}
}

\begin{document}
%
\title{Probabilistic Semantic Retrieval for Surveillance Videos with Activity Graphs}
%
%
%

\author{Yuting~Chen,
        Joseph~Wang,
        Yannan~Bai,
        Gregory~Castañón,
        and~Venkatesh~Saligrama
\thanks{Y. Chen, Y. Bai and V. Saligrama are with Boston University, Boston,
MA 02215 USA (e-mail: yutingch@bu.edu, ynbai@bu.edu, srv@bu.edu).}
\thanks{J. Wang is with Amazon, Cambridge, MA, 02142 (e-mail: wangjose@amazon.com). This work was done while he was with Boston University.}
\thanks{G. Castañón is with Systems and Technology Research, Woburn, MA, 01801 (e-mail:gregory.castanon@stresearch.com). This work was done while he was with Boston University.}}

%
%

\markboth{IEEE TRANSACTIONS ON MULTIMEDIA}
{Chen \MakeLowercase{\textit{et al.}}: Probabilistic Semantic Retrieval for Surveillance Videos with Activity Graphs}
%




\maketitle
\IEEEpeerreviewmaketitle
\begin{abstract}
We present a novel framework for finding complex activities matching user-described queries in cluttered surveillance videos. 
The wide diversity of queries coupled with unavailability of annotated activity data limits our ability to train activity models.
To bridge the semantic gap we propose to let users describe an activity as a semantic graph with object attributes and inter-object relationships associated with nodes and edges, respectively. We learn node/edge-level visual predictors during training and, at test-time, propose to retrieve activity by identifying likely locations that match the semantic graph.
We formulate a novel CRF based probabilistic activity localization objective that accounts for mis-detections, mis-classifications and track-losses, and outputs a likelihood score for a candidate grounded location of the query in the video. We seek groundings that maximize overall precision and recall. To handle the combinatorial search over all high-probability groundings, we propose a highest precision subgraph matching algorithm.  Our method outperforms existing retrieval methods on benchmarked datasets.    
\end{abstract}


%
\IEEEpeerreviewmaketitle

\section{Introduction}
\label{sec:search_introduction}
The rapid growth of surveillance camera networks over the last decade has created a critical need for autonomous video analysis systems that can reason over large video corpora. Many routine tasks such as activity detection, anomaly detection and activity recognition \& retrieval in surveillance videos currently require significant human attention. The goal of this paper is to develop exploratory search tools for rapid analysis by human operators.

Video retrieval aims to recover spatial-temporal locations of topics of interest from a large video corpora. Unlike typical approaches \cite{Yu2016retrieval,Liong2017deepvideo,Lin2017deepretrieval} that require exemplar videos, we focus on retrieval of activity that matches a user's description, or {\it {\bf a}nalyst or user {\bf d}escribed {\bf s}emantic {\bf a}ctivity (ADSA)} query, from surveillance videos. Surveillance videos pose two unique issues: (a) wide query diversity; (b) the presence of many unrelated, co-occurring activities that share common components.

\begin{figure}
\centering
    \begin{subfigure}[b]{0.23\textwidth}
        \includegraphics[trim=60mm 100mm 60mm 110mm,clip,width=\textwidth]{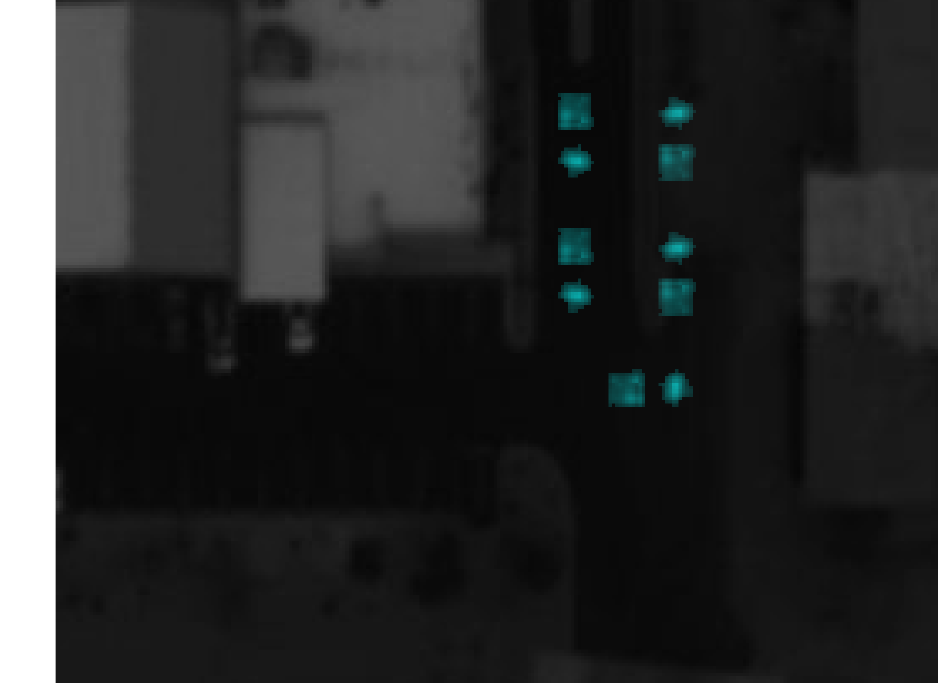}
        \caption{U-Turn}
        \label{fig:uturn}
    \end{subfigure}
    \begin{subfigure}[b]{0.23\textwidth}
        \includegraphics[trim=60mm 100mm 60mm 110mm,clip,width=\textwidth]{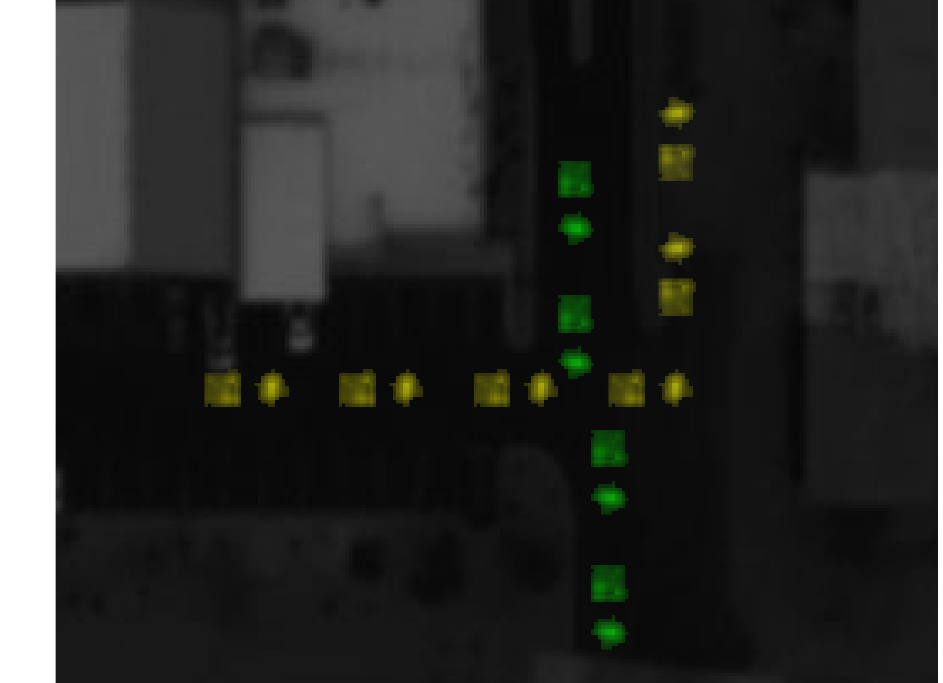}
        \caption{False Alarm}
        \label{fig:false_uturn}
    \end{subfigure}
    \caption{An example of the need for relationships between components of an action, in this case retrieving a u-turn in a wide-area-motion-imagery (WAMI) data. Even for this simple, single object activity, relationships between detections are important to define the activity. Ignoring relationships between detections in (b), notably that the perceived components of the ``u-turn'' are due to two different vehicles, yields a false alarm.}\label{fig:introfig}
\end{figure}

\noindent

\begin{figure*}[t]
\begin{center}
\includegraphics[width=5.2in]{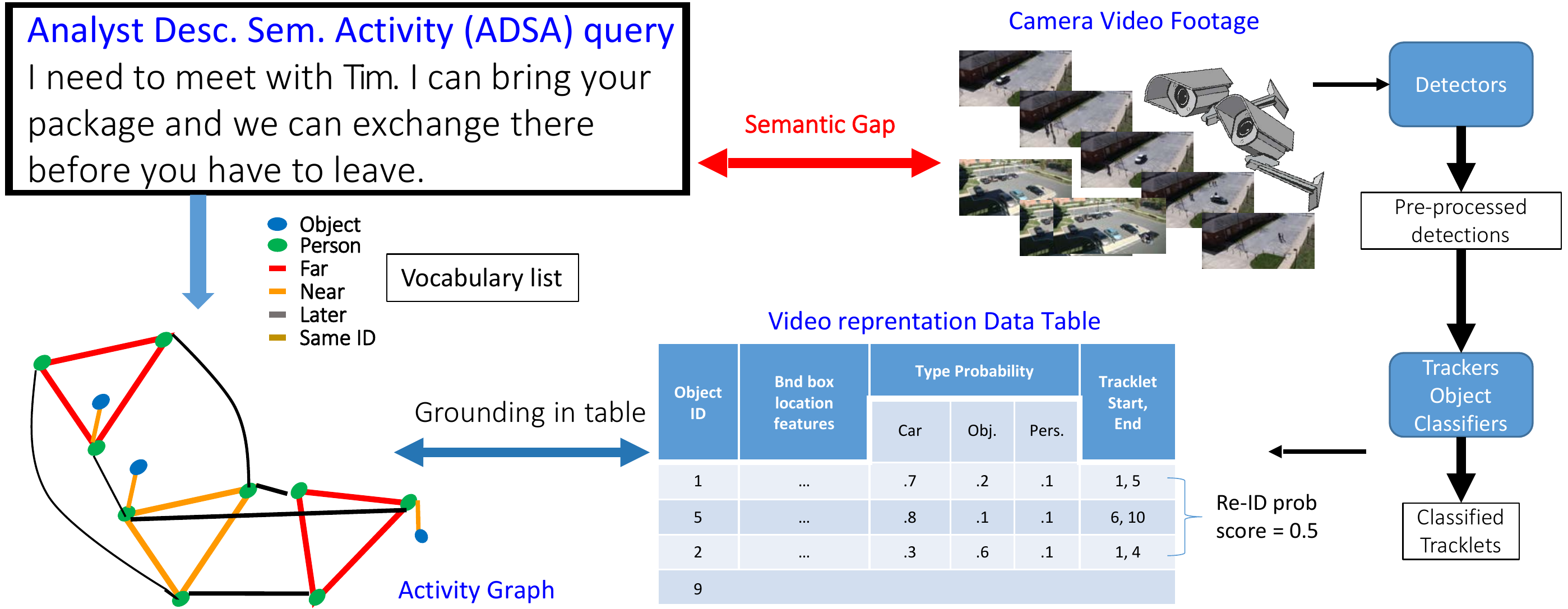}
\caption{Overview of Proposed Probabilistic Semantic Retrieval Approach (see Sec.~\ref{sec:search_introduction} and Sec.~\ref{sec:search_implementation})}
\label{fig:overview}
\end{center}
\end{figure*}

The wide diversity of ADSAs limits our ability to collect sufficient training data for different activities and learn activity models for a complete list of ADSAs. Methods that can transfer knowledge from detailed activity descriptions to the visual domain are required. As noted in \cite{Johnson}, while it would be desirable to learn to map textual descriptions to a semantic graph, this by itself is an active area of research. To handle query diversity, we focus on a novel intermediate approach, wherein a user represents an activity as a semantic graph (see Fig.~\ref{fig:overview}) with object attributes and inter-object semantic relationships associated with nodes and edges respectively. We propose to bridge the relationship semantic gap by learning relationship concepts with annotated data.
At the object/node-level, we utilize existing state-of-art methods to train detectors, classifiers and trackers to obtain detected outputs, class labels, track data and other low-level outputs. This approach is practical because, in surveillance, the vocabulary of low-level components of a query is typically limited and can be assumed to be known in advance. 

Our next challenge is to identify candidate groundings. By a \textit{grounding} \cite{Johnson}, we mean a mapping from archive video spatio-temporal locations to query nodes (see also Sec.~\ref{sec:activity_models}). 
Finding groundings of a query in a video is a combinatorial problem that requires searching over different candidate patterns that matches the query.
The difficulty arises from many unrelated co-occurring activities that share node and edge attributes. Additionally, the outputs of low-level detectors, classifiers and trackers are inevitably error-prone leading to mis-detections, mis-classifications, and loss of tracks. Uncertainties can also arise due to the semantic gap.
Consequently, efficient methods that match the activity graph with high-confidence in the face of uncertainty are required. 

This paper extends our preliminary work on activity retrieval \cite{Castanon_mm15} with a novel probabilistic framework to score the likelihood of groundings and explicitly account for visual-domain errors and uncertainties. \cite{Castanon_mm15} proposes to identify likely candidate groundings as a ranked subgraph matching problem. By leveraging the fact that the attributes and relationships in the query have different level of discriminability, a novel maximally discriminative spanning tree (MDST) is generated as a relaxation of the actual activity graph to quickly minimize the number of possible matches to the query while guaranteeing the desired recall rate. In \cite{Castanon_mm15}, the activity graph that describes semantic activity requires a fixed and manual description of node attributes and edge relationships, which relies heavily on domain knowledge and is prone to noise from lower-level pre-processing algorithms. 

In this paper, we propose a probabilistic framework based on a CRF model of semantic activity that combines the activity graph with the confidence/margin outputs of our learned component-level classifiers, and outputs a likelihood score for each candidate grounding. 
We pose the combinatorial problem of identifying likely candidate groundings as a constrained optimization problem of maximizing precision at a desired recall rate. To solve this problem we propose a successive refinement scheme that recursively attempts to find candidate matches at different levels of confidence. For a given level of confidence, we show that a two-step approach based on first finding subgraphs of the activity graph that are guaranteed to have high precision, followed by a tree-based dynamic programming recursion to find the matches, leads to efficient solutions. 
Our method outperforms bag of objects/attributes approaches \cite{Lin_2014_CVPR}, demonstrating that objects/attributes are weak signatures for activity in surveillance videos unlike other cases \cite{Xu_2015_CVPR,Jain_2015_CVPR,karpathy_2014_CVPR,Simonyan_NIPS_2014}. We compare against approaches \cite{Castanon_mm15} based on manually encoding node/edge level relationships to bridge the visual domain gap and demonstrate that our semantic learning combined with probabilistic matching outperforms such methods.

\subsection{Related Work}
\label{ssec:related_work}
In the context of multimedia video analysis, several areas are related to the proposed approach. 

{\it Action Recognition \& Event Detection Methods:} Many video retrieval methods solve the retrieval problem by classification  \cite{tang2012learning,yang2013related,ma2014knowledge,Lin_2014_CVPR,Shu_2015_CVPR,won14,wang2014event}, namely, at run-time they take in a video snippet (temporal video segment) as input and output a score based on how well it matches the desired activity. During training, activity classifiers for video snippets are learned using fully labeled training data. In this context, several recent works have proposed deep neural network approaches to learn representations for actions and events~\cite{Xu_2015_CVPR,Jain_2015_CVPR,karpathy_2014_CVPR,Ye2015deepvideo,Liong2017deepvideo}. These works leverage the fact that in some applications object/attributes provide good visual signatures for characterizing activity. 

There have been a number of works that have been proposed recently in this context. Zhao et al. \cite{zhao2017videowhisper} learn video representations from video sequences using attention-based recurrent neural networks. Zhang et al. \cite{zhang2018discriminative} propose a method that account for correlation among parts followed by refinement of the candidate space. They show that their method leads to significant improvement in action recognition accuracy. Wang et al. \cite{wang2018two} propose a novel LSTM/CNN-E model to learn a global description for the input video using time-varying descriptions extracted from the STPP ConvNet. Kumar et al. \cite{kumar2018f} propose a local-alignment-based FASTA based approach to summarize events in multi-view videos. 

Nevertheless, in contrast to these methods, we do not utilize any activity-level training data. Furthermore, while these methods are suited for situations where an activity manifests as a dominant signature in the video snippet, they are ill-suited for situations where the activity signature is weak, namely, the activity occurs among many other unrelated co-occurring activities, which is the typical scenario in surveillance problems.   

{\it Multimedia Video Representation:}
In multimedia video recognition and retrieval, a number of works leverage information from multiple modalities or sources such as text, audio, lower level visual content, OCR and higher level visual concepts in combination with video \cite{Hoi2008Multi,Xu2008Multi,Chen2012Multi,Merler2012event,Ma2013event}. 
Merler et al. \cite{Merler2012event} and Ma et al. \cite{Ma2013event} utilize external images and videos to build an intermediate level video representation for event detection. Mazloom et al. \cite{Mazloom2016event} learn a video descriptor based on the tags of their nearest neighbors in a large collection of social tagged videos. Song et al. \cite{Song2017event} extract key segments for event detection by transferring concept knowledge from web images and videos. Xu et al. \cite{Xu2008Multi} target the sport video event detection problem by aligning broadcast video and accompanied text. Chen et al. \cite{Chen2012Multi} present a mutual information variant to fuse features from audio and video domain to perform action recognition and retrieval. Pang et al. \cite{Pang2015retrieval} adopt the deep Bolzmann machine (DBM) to build a joint density model over video, audio and text domains for emotion classification and cross-modal retrieval. \cite{shi2017sequential} propose to project dense trajectories into two-dimensional planes, and subsequently a CNN-RNN network is employed to learn an effective representation for long-term motion.
Our current implementation does not rely on any such multimedia information, and can be easily extended by adding attributes and relationships in corresponding domains when necessary.

{\it Video Hashing Methods:}
Hashing based methods focus on the computational needs of large-scale video retrieval by comparing a compact representation of query and archive videos learned from unsupervised or supervised data \cite{Yu2016retrieval,Hao2017hashing,Liong2017deepvideo}. Yu et al. \cite{Yu2016retrieval} learn a hashing model based on extracting key frames and imposing pairwise constraints for semantically similar frames. Hao et al. \cite{Hao2017hashing} propose an unsupervised hashing algorithm that combines multiple feature representations from key frames. Liong et al. \cite{Liong2017deepvideo} learn binary codes for the entire video with a deep learning framework without the help of individual key frames. In contrast, our method applies to scenarios where exemplar videos are not available, and queries are instead described by users in terms of activity graphs.

{\it Zero-shot Methods:} 
More recently, zero-shot methods have been applied to several visual tasks such as video retrieval\cite{han2017vrfp}, event detection \cite{wu2014zero,chang2015semantic,elhoseiny2015zero}, action recognition \cite{gan2015exploring}, action localization~\cite{Jain_2015_ICCV}, image tagging \cite{Zhang_2016_CVPR}, and image recognition~\cite{lampert2009attribute, wang2016zeroshot-reid}. These methods share the same advantage with our work in that activity level training data associated with the desired activity is not required. 
Nevertheless, zero-shot methods are trained based on source domain descriptions for a subset of activities that allow for forging links between activity components, which can then be leveraged for classification of unseen activity at test-time. 
Furthermore, the current set of approaches are only suitable in scenarios where the activity has strong visual signatures in low-clutter environments. 

Other recent methods such as  \cite{han2017vrfp} describe on-the-fly retrieval based on obtaining training data after the query is input (e.g., from the web). These methods although different from zero-shot methods still require training data that our methods does not.

{\it Activity Graphs:}
It is worth pointing out that several works  \cite{Lin_2014_CVPR,Shu_2015_CVPR,won14,Choe_2013_ICCV} have developed structured activity representations but they use fully annotated data as mentioned earlier. Lin et al. \cite{Lin_2014_CVPR} describe a bipartite object/attribute matching method. Shu et al. \cite{Shu_2015_CVPR} describe AND-OR-Graphs based on aggregating sub-events for test-time activity recognition. Similar to classification based methods, these approaches only work well when the desired activity is dominant over a video snippet. 

The proposed method is closely related to our preliminary work \cite{Castanon_mm15}. Activities are manually represented as graph queries. Ground-truth data is utilized to reduce video to a large annotated graph. A ranked subgraph matching algorithm is used to find matches in the video archive graph. In this way object-level semantic gap is avoided. Relationship semantic gap is handled manually (for instance, nearness, proximity etc are entered manually in terms of pixel distances). This is somewhat cumbersome because relationships are often context dependent. It is primarily a deterministic subgraph matching solution that does not handle visual distortion like mis-detections and tracker failure well.
In contrast we formulate a probabilistic activity graph that explicitly accounts for visual distortions, bridges the semantic gap through learning low-level concepts, and proposes an efficient probabilistic scoring scheme based on CRFs.

{\it CRF Models for Retrieval:}
Our proposed CRF framework closely resembles CRF models that are employed for semantic image retrieval in Johnson et al. \cite{Johnson,zitnick}. They propose scene graphs to represent objects and relationships between them, and train a CRF model using fully annotated training data. These CRF models on fully trained data thus can also incorporate knowledge of typical global scenes and context in addition to low-level node/edge predictions. 
In contrast our premise is that, in the video problem, we do not have adequate training data across all desired activities. In addition, unlike images, miss detections and track losses have substantial impact in video retrieval. 
Finally, spatio-temporal scale and size of the surveillance videos, and the presence of unrelated co-occurring activities lead to a probabilistic and combinatorial search problem.   

\section{Activity Models}
\label{sec:activity_models}
The goal of semantic activity retrieval is to spatio-temporally ground semantically described activities in large videos. As no examples of activities are provided, a semantic framework is necessary to represent the search activity. To capture activities involving multiple objects over potentially large temporal scale, we need a flexible framework capable of representing both the objects involved in the activity as well as relationships between these objects. To capture both the components of the activity as well as their relationships, we use an \textit{activity graph} to define a query. 

An activity graph is an abstract model for representing a user-described activity that captures object entities, their attributes, and spatio-temporal relationships between objects. An activity graph provides a detailed description of video activity because it admits diverse sets of objects, attributes, and relationships. Graphs represent a natural approach to representing activities involving interaction between multiple objects. For example, consider the following activity:

\noindent\textit{Two men are meeting so one can give the other a backpack. They will meet and talk first, then they will go to a vehicle and drive away. One man is wearing a red shirt, the other is wearing a green shirt, and their vehicle is a blue sedan.}

The above description can be represented as a composition of atomic elements, element descriptions, relationships between elements, and relationship descriptions. For example, the activity can be described by 4 atomic elements with specific descriptions, a person wearing red (P1), a person wearing green (P2), an object representing a backpack (O), and a blue car (V). Using these elements, the activity can be described by the interactions between these elements: initially, P1 and O are near each other, then P1, P2, and O are near each other. The three objects P1, P2, and O move near V, then O, P1, and P2 enter V, and finally, V moves.

Formally, an \textit{activity graph} is composed of nodes, each representing a realization of an object at an instance of time, and edges, representing relationships between nodes. 

We adapt the notation used in scene recognition \cite{Johnson} and assume we are given a set of object classes $\mathcal{C}$, a set of attributes $\mathcal{A}$ associated with each object, and a set of relationships $\mathcal{R}$ between objects.

An \textit{activity graph} $G$ is defined as the tuple $G=(O,E)$. $O$ denotes the $n$ nodes in the graph, $O=\{o_1,\ldots,o_n\}$, with each node characterized by its class and attributes, $o_i=(c_i,A_i)\in \mathcal{C}\times\mathcal{A}$. Similarly, $E \subseteq |O|\times |O| \times \mathcal{R}$ denotes the set of edges between nodes of the graph, with each edge $e_{ij}=(o_i,R_{ij},o_j)$ characterized by its associated relationships $R_{ij}=\{r^{(1)}_{ij},\ldots,r^{(n)}_{ij}\}$, where $R_{ij} \subseteq \mathcal{R}$ represents the set of relationships between objects $o_i$ and $o_j$.

Differing from image retrieval, edges in an activity graph represent not only spatial displacement, but additionally temporal displacement as well as identity information, to capture concepts such as ``\textit{the same person is near the vehicle later}.'' Similarly, attributes associated with nodes also include time-dependent attributes such as velocity.

In searching for activities, we seek to \textit{ground} the activity graph to a video segment, that is to associate each node and edge in our activity graph with parts of the video denoted by spatio-temporal bounding boxes $B$. For the nodes of an activity graph, $O$, and a set of bounding boxes $B$, a \textit{grounding} $\gamma:O\rightarrow B$ is a mapping between nodes and bounding boxes. Note that mapping nodes to bounding boxes is sufficient to map the graph $G$ to the video segment as the edges are implicitly mapped by $\gamma$. For a grounding $\gamma$, we denote the bounding box that element $o_i$ is mapped to by $\gamma$ as $\gamma_i$.

In this framework, the problem of semantic activity retrieval is equivalent to choosing a grounding for the activity graph. In Section \ref{sec:graph_retrieval}, we formulate an approach to efficiently grounding an activity graph in a large archive video.

Representing text as an activity graph requires mapping of nouns to objects and understanding of relationships, activities, and interaction between elements. This work is out of the scope of this paper, and to better demonstrate the efficacy of our approach to retrieval, we focus solely on the problem of spatio-temporally locating activities in videos given a human-generated activity graph. In practice, these activity graphs are composed of components that are semantically interpretable to humans.

\section{Activity Retrieval by Graph Grounding}\label{sec:graph_retrieval}
Our goal is to find an activity in a large archive video. To this end, we seek to find a grounding of an activity graph, representing the query activity, in the archive video. To solve this problem, we must address two main sub-problems: how to evaluate the grounding between an activity graph and a set of object bounding boxes (generated from object proposal approaches like~\cite{cheng2014bing}), and how to search over a large archive of bounding boxes in order to find the highest scoring grounding. We first present an approach to evaluate a grounding between activity graph and bounding boxes, then present an approach to efficiently reason over a large archive video to finding the optimal groundings.

\subsection{Evaluating Activity Graph Grounding}
To evaluate the grounding between an activity graph and set of bounding boxes, we consider a maximum a posteriori inference scheme. For a graph $G=(O,E)$, set of bounding boxes $B$, and grounding $\gamma$, we consider the maximum a posteriori probability, that is $P(\gamma|G,B)$. We consider a conditional random field (CRF) model \cite{Lafferty:2001:CRF:645530.655813},
\begin{align}\label{eqn.grounding_model}
P(\gamma|G,B)=\prod_{o \in O}P(\gamma_o|o)\prod_{o,r,o'\in E}P(\gamma_o,\gamma_{o'}|o,r,o').
\end{align}
Given that we are in a zero-shot setting, we consider uniform distributions over bounding boxes, $p(\gamma_o)$, and activity graph nodes, $p(o)$. From Bayes' rule, the conditional probability can be expressed
\begin{align*}
P(\gamma|G,B)=\prod_{o \in O}P(o|\gamma_o)\frac{p(\gamma_o)}{p(o)}\prod_{o,r,o'\in E}P(\gamma_o,\gamma_{o'}|o,r,o').
\end{align*}
Our goal is to find the maximum a posteriori grounding,
\begin{align}\label{eqn.MAP_grounding}
\max_{\gamma}\prod_{o \in O}P(o|\gamma_o)\prod_{o,r,o'\in E}P(\gamma_o,\gamma_{o'}|o,r,o').
\end{align}
Note that due to the uniform distribution assumptions on $p(o)$ and $p(\gamma_o)$, these terms are constant and are ignored in finding the maximum a posteriori grounding.

\subsubsection{Learning Node \& Edge Level Probability Models} \label{sssec:node_edge_model}

To evaluate the maximum a posteriori probability of a grounding, the distributions $P(o|\gamma_o)$ and $P(\gamma_o,\gamma_{o'}|o,r,o')$ need to be estimated. The distribution $P(o|\gamma_o)$, representing the probability that the bounding box specified by $\gamma_o$ has the class, $c$, and attributes $a$, associated with node $o$. We assume that the probabilities of class and attributes are independent, and therefore we can model this as a product of distributions:
\begin{align}
P(o|\gamma_o)=P(c|\gamma_o)\prod_{a \in A}P(a|\gamma_o).
\end{align}
Estimating each of these probabilities is accomplished by learning an object detector or attribute classifier, with the output margin mapped to a probability using a logistic model. 
\begin{align}
P(c|\gamma_o) =   \dfrac{1}{1+\exp{(sf_c(\gamma_o)+t)}} \nonumber
\end{align}
where $f_c(\cdot)$ is the output margin of the detector for class $c$, $s$ and $t$ are two scalar parameters that can be set heuristically or learned with Platt scaling \cite{platt1999probabilistic}.

Similarly, we learn semantic relationship classifiers on features from detected object pairs, and estimate the distribution $P(\gamma_o,\gamma_{o'}|o,r,o')$ as in the case of object probabilities. 
Our perspective is that the vocabulary typically used to describe complex activity by analysts is a priori known (''carry, together, with, near''). In \cite{Castanon_mm15}, relationship was manually annotated as it is convenient for the subsequent matching stage since everything was deterministic. However it limits the method to work robustly in different scenarios.
For example, consider the relationship \textit{near} between two objects. The manually way is to set a pixel-distance threshold for identifying ``near'' property for two objects. However, the semantic meaning of \textit{near} is strongly dependent on the context of the two objects. \textit{Near} in the context of moving vehicles is different from stationary vehicles or for two persons.

The details of the node and edge probability models we used are described in Sec. \ref{sec:search_implementation}.

\subsection{Efficient Grounding in Large Graphs}
In the previous section, we presented an approach to estimate the conditional probability of a grounding for a given activity graph. Although estimating the probability of a specific grounding can be efficiently achieved, a combinatorially large number of possible groundings exist between an activity graph and collection of bounding boxes. Furthermore, due to long surveillance videos, the collection of extracted bounding boxes is generally large.

In order to efficiently find the maximum a posteriori grounding of an activity graph in a video, we instead consider the following optimization problem:
\begin{align}\label{eqn.thresh_grounding}
\max_{\gamma}\prod_{o \in O}\mathds{1}_{P(o|\gamma_o)\geq \tau_o}\prod_{o,r,o'\in E}\mathds{1}_{P(\gamma_o,\gamma_{o'}|o,r,o')\geq \tau_{o,o'}}.
\end{align}
Note that for the proper setting of thresholds $\tau_o$ and $\tau_{o,o'}$, the solution of \eqref{eqn.thresh_grounding} is equivalent to the solution of \eqref{eqn.MAP_grounding}. In the case where the parameters are set below this optimal set of parameters, the solution is non-unique, with a set of possible groundings returned, one of which is the solution to \eqref{eqn.MAP_grounding}. By scoring the groundings that maximize \eqref{eqn.thresh_grounding} according to the objective of \eqref{eqn.MAP_grounding}, we are able to find the optimal grounding from this subset.

Our goal is to find a set of thresholds $\tau$ that maximize precision subject to a recall constraint. For a grounding $\gamma$, we define $F$ as the value of the objective of \eqref{eqn.thresh_grounding}, that is
\begin{align*}
F(\gamma,\tau)=\prod_{o \in O}\mathds{1}_{P(o|\gamma_o)\geq \tau_o}\prod_{o,r,o'\in E}\mathds{1}_{P(\gamma_o,\gamma_{o'}|o,r,o')\geq \tau_{o,o'}}.
\end{align*}
Let $y_{\gamma}$ denote whether or not a grounding $\gamma$ corresponds to the desired activity, for a set of thresholds $\tau$, the precision can be expressed as the probability of a grounding corresponding to the desired activity having an objective value, $F$, equal to $1$ divided by the probability of any grounding having an objective value equal to $1$, that is: 
\begin{align*}
\mbox{Prec}(\tau)=\frac{P\left(F(\gamma,\tau)=1|y_{\gamma}=1\right)P(y_{\gamma}=1)}{P\left(F(\gamma,\tau)=1\right)},
\end{align*}
We assume the probability of a grounding corresponding to the desired activity is significantly smaller than the probability of a grounding not corresponding to the desired activity, allowing for the approximation:
\begin{align*}
P\left(F(\gamma,\tau)=1\right)\approx P\left(F(\gamma,\tau)=1|y_{\gamma}=0\right)P(y_{\gamma}=0).
\end{align*}
Similarly, we can express the recall rate as 
\begin{align*}
\mbox{Rec}(\tau)=P(F(\gamma,\tau)=1|y_{\gamma}=1).
\end{align*}
We therefore seek to minimize the approximate precision subject to the recall rate being greater than some value $\eta$:
\begin{align*}
\min_{\tau}&\frac{P\left(F(\gamma,\tau)=1|y_{\gamma}=1\right)P(y_{\gamma}=1)}{P\left(F(\gamma,\tau)=1|y_{\gamma}=0\right)P(y_{\gamma}=0)}.\\
&\mbox{s.t.  } P(F(\gamma,\tau)=1|y_{\gamma}=1)\geq \eta \end{align*}
Note that the ratio $\frac{P(y_{\gamma}=1)}{P(y_{\gamma}=0)}$ is an unknown quantity dependent on the archive video, however as this quantity is a constant, the value does not effect the optimization. Assuming independence of the nodes and edges of the activity graph (and their attributes), the remaining conditional properties can be estimated by evaluating detector performance, with thresholds chosen given detector performance.

Solving the optimization problem in \eqref{eqn.thresh_grounding} has the potential to be significantly more efficient than solving the optimization problem in \eqref{eqn.MAP_grounding} through the use of branch-and-bound approaches, particularly due to the ability to aggressively bound by eliminating any solution where one node or edge does not meet the associated threshold. Unfortunately, despite the potential improvement in efficiency, solving this problem is still combinatorially hard and may be computationally infeasible, particularly for a large collection of bounding boxes. 

Rather than directly solving this problem, consider a subgraph of $G$ that we denote as $\hat{G}\subseteq G$, with the nodes and edges of the subgraph denoted $\hat{G}=\left(\hat{O},\hat{E}\right)$. Consider the problem of finding a grounding for this subgraph:
\begin{align}\label{eqn.subgraph_thresh_grounding}
\max_{\gamma}\prod_{o \in \hat{O}}\mathds{1}_{P(o|\gamma_o)\geq \tau_o}\prod_{o,r,o'\in \hat{E}}\mathds{1}_{P(\gamma_o,\gamma_{o'}|o,r,o')\geq \tau_{o,o'}}.
\end{align}
For this subgraph matching problem, we make the following observation:
\begin{thm}\label{thm.subgraph_recall}
Any grounding of the graph $G$ that maximizes \eqref{eqn.thresh_grounding} is also a subgraph grounding that maximizes \eqref{eqn.subgraph_thresh_grounding}.
\end{thm}
Thm. \ref{thm.subgraph_recall} implies that the set of groundings that maximize \eqref{eqn.subgraph_thresh_grounding} includes all groundings that also maximize \eqref{eqn.thresh_grounding}. Therefore, the set of groundings that maximize \eqref{eqn.subgraph_thresh_grounding} has a recall rate of $1$, though the precision rate may be decreased.

Thm. \ref{thm.subgraph_recall} leads to an efficient approach to solving \eqref{eqn.thresh_grounding}. Rather than searching for the full graph $G$, we instead consider a subgraph $\hat{G}$ that can be efficiently searched for. From Thm. \ref{thm.subgraph_recall}, all subgraphs of $G$ will have a recall rate of $1$, however the choice of spanning tree directly impacts the precision rate of the set of groundings that maximize \eqref{eqn.subgraph_thresh_grounding}. We therefore propose selecting a Highest Precision Subgraph (HPS) defined as the subgraph of $G$ with a minimal expected number of groundings that maximize \eqref{eqn.subgraph_thresh_grounding}.

In particular, we attempt to find a HPS $\hat{G}$ from the set of spanning trees of $G$, as tree search can be efficiently solved using dynamic programming \cite{castanon2012exploratory}. From our model in \eqref{eqn.grounding_model}, we assume that each edge is distributed independently. Therefore, we can find the HPS from the set of spanning trees of $G$ by finding a spanning tree over the graph $G$ that minimizes the likelihood that an edge probability is above the associated threshold.
\begin{align}\label{eqn.hps_tree}
\argmin_{\hat{E}}\smashoperator{\sum_{o,r,o'\in \hat{E}}}\log\left(\mathds{E}_{o,r,o'\sim \mathcal{\psi}}\left[\mathds{1}_{P(\gamma_o,\gamma_{o'}|o,r,o')\geq \tau_{o,o'}}\right]\right),
\end{align}
where $\hat{E}$ is restricted to be the set of edges that yield a valid spanning tree over $G$ and $\mathcal{\psi}$ is the distribution over relationships in the video. In practice, $\mathcal{\psi}$ can be efficiently approximated by randomly sampling bounding box pairs and estimating their distribution. Solving the optimization in \eqref{eqn.hps_tree} can be done efficiently, as the problem can be mapped to a minimum spanning tree problem. We explain the details of the algorithm in Sec. \ref{sec:search_algorithm}.

\section{Highest Precision Subgraph Matching Algorithm}

\label{sec:search_algorithm}
The goal of our algorithm is to efficiently retrieve the optimal grounding $\gamma$ for an activity graph $G$. We accomplish it in two steps: First, we calculate the highest precision subgraph $\hat{G}$ for an activity graph $G$. The selected subgraph minimizes \eqref{eqn.hps_tree} among all the spanning trees of $G$, thus filters out as many infeasible groundings as possible. Then we develop a Highest Precision Subgraph Matching (HPSM) approach to find the optimal groundings that maximizes \eqref{eqn.subgraph_thresh_grounding}. In the end, we recover the ranked groundings with respect to the original activity graph $G$.

\subsection{Highest Precision Subgraph Selection}
\label{sec:tree-selection}
Given an activity graph $G=(O,E)$, we first reduce the video data to the set of potentially relevant nodes and edges by building a coarse archive graph $C=G(O^c,E^c)$ out of the spatio-temporal bounding boxes $B$.
For every node $o = (c, A) \in O$, we retrieve the set of corresponding locations $o^c_{i}$ that satisfy the class and attributes characterized by $c$ and $A$. 
Similarly, we retrieve the corresponding edges $e^c_{ij}=(o^c_i,o^c_j,R_{ij})$. 
    
Despite incredible cost savings, the down-sampled coarse graph is still a large graph with a collection of hundreds of thousands of bounding boxes. We therefore select a HPS $\hat{G}$ from spanning trees of the activity graph $G$ so as to minimize the time spent performing an expensive search. 
    
The choice of which spanning tree to select has significant run-time implications. The creation of a spanning tree involves the removal of edges from $G$, and not all edges are created equal. The edges in the HPS $\hat{G}$ should be the set of edges that minimizes the likelihood in \eqref{eqn.hps_tree}. To solve the optimization, we first compute a set of weights indicating the discriminative power of the edges, then calculate the spanning tree $T$ which minimizes the total edge weight.

\subsubsection{Weight Computation}
\label{sec:weights}
During the archival process, we assign probabilities $p(c)$, $p(a)$ and $p(r)$ to each class, attribute and relationship that we store. These functions denote the probability that a randomly-chosen class or attribute or relationship in the archive video is a match to the class $c$, attribute $a$ or relationship $r$. Relationships, in particular, have greater power to be discriminative because the set of potential relationships is $|B|^2$. 

The set of edges that minimizes \eqref{eqn.hps_tree} is associated with the most discriminative relationships so that it yields the least possible mappings in the coarse archive graph $C$. In VIRAT\cite{oh2011large} datasets, while the ``Person near car'' relationship is normally very discriminative, 80\% of the dataset is shot in parking lots, where people are frequently near cars.  Objects disappear near cars far less less frequently - thus, a tree rooted at the ``object disappears'' node and connecting through the ``near'' edge to the ``car'' node has less potential matches than one starting elsewhere.

We compute empirical values for $p(r)$ during the archival process by computing the percentage of  relationships that has appeared in the videos.  If relationships have not been seen, they are assumed to be nondiscriminative and assigned values of $p(r)=1$.  If it is later determined that these relationships are discriminative, we can revise our estimate of $p(r)$.

\subsubsection{Highest Precision Spanning Tree}
From our model in \eqref{eqn.grounding_model}, We assume that each edge is distributed independently. Absent additional information indicating the distribution of relationships in the video corpus, we assume that all  relationships are generated independently. 

Since relationships and edges in an activity graph $G$ = ($O$, $E$) are independent, the total edge weight of the graph is:

\begin{align}
\displaystyle p(E) = \prod_{e_{ij} \in E}\prod_{r^{(k)}_{ij} \in R_{ij}}p(r^{(k)}_{ij} )
\end{align}

As noted in \eqref{eqn.subgraph_thresh_grounding}, we are going to do a search using HPS $\hat{G}$ instead of the original query graph $G$ in order to reduce the computational complexity. That HPS $\hat{G}$ which results in the fewest possible groundings is our novel highest precision spanning tree.

\begin{definition*}[Highest Precision Spanning Tree (HPST)] 
We call a spanning tree $T^*$ an HPST of activity graph $G$ with  edge weights $p(e) \forall e\in E$, if the tree satisfies
\begin{align}
\label{eqn.mdst}
T^*=\argmin_{T(\hat{O},\hat{E})\in\mathcal{T}}p(\hat{E}) = \argmin_{T(\hat{O},\hat{E})\in\mathcal{T}}\sum_{e_{ij}\in \hat{E}}\log p(e_{ij}),
\end{align}
where $\mathcal{T}$ denotes the set of all possible spanning trees induced from activity graph $G$.
\end{definition*}

This is exactly the same as the definition of minimum spanning tree. By minimizing the total edge weight, we achieve a Highest Precision Spanning Tree $T^*$, which solves the optimization in \eqref{eqn.hps_tree}. We use Kruskal's algorithm to calculate the HPST that should produce the fewest possible matches.

\subsection{Highest Precision Subgraph Matching (HPSM)}
\label{sec:optimization}
Given a coarse graph $C$ and the HPST $T^*$, we seek to select the maximum a posteriori grounding $\gamma: O^* \to B$ from all possible grounding $\gamma's$. We solve for the optimal grounding between the HPST $T^*$ and archive graph $C$ in two steps. In the first step, we construct a \emph{matching graph} $H$ of the possible groundings. Then we find the optimal grounding from the matching graph $H$.

\subsubsection{Matching Graph Creation}
 we build a matching graph $H=G(O^h,E^h)$, where each node $o \in O^h \subseteq O^* \times O^c $ is a tuple denoting a proposed assignment between a node in $T^*$ and a node in $C$, and each edge $e \in E^h$ denotes the relationship between the two assignments. All the assignments in $H$ satisfy the setting of nodes and edges thresholds $\tau$ described in \eqref{eqn.subgraph_thresh_grounding} so that we can rule out the impossible mappings.

We create $H$ by first adding assignments for the root, then adding in assignments to its successors which satisfy both node and edge relationships described in $T^*$. We then set the score thresholds $\tau_o$ and $\tau_{o,o'}$ to be the minimum score for nodes and edges and find a set of mappings that maximize \eqref{eqn.subgraph_thresh_grounding}. The proper setting of thresholds ensures that no feasible groundings to $T^*$ is ruled out in the filtering process. This process is described in Algorithm \ref{alg:matching-graph}, and the expected number of mappings scales as a product of $p(E^*)$ and the size of the archive data.

\begin{algorithm}[ht!]
  \caption{Create Matching Graph}
  \label{alg:matching-graph}
  \begin{algorithmic}[1]
    \Procedure{Create Matching Graph}{$T^*, C, P(o|\gamma_o), P(\gamma_o,\gamma_{o'}|o,r,o')$}
      \State{$H=G(O^h,E^h) \gets \emptyset$}
      \State{Iterate from root to leaves}
      \ForAll{$o^* \in O^*$}
      	\State{Compute the groundings to this node}
      	\State{$N_{o^*} \gets (o^*,o^c)$ where $P(o^*|\gamma_{o^*}=o^c)>\tau_o$}
      	\If{$Parent(o^*)\neq \emptyset$}
      		\State{$E \gets \emptyset$}
      		\ForAll{$(o^*_p,o^c_p) \in N_{Parent(o^*)}$}
      			\State{$E \gets E \cup (o^*,o^c)$ where $P(\gamma_{o^*}=o^c,\gamma_{o^*_p}=o^c_p|o^*,r,o^*_p)>\tau_{o,o'}$}
      		\EndFor
      		\State{$N_{o^*} \gets N_{o^*} \cap E$}
      		\State{$O^h \gets E^h \cup N_{o^*}$}
      		\State{$E^h \gets E^h \cup E$}
      	\Else
      		\State{$O^h \gets O^h \cup N_{o^*}$}
      	\EndIf
			\EndFor
    \EndProcedure
  \end{algorithmic}
\end{algorithm}

\subsubsection{Retrieval with HPSM}
After traversing $T^*$ from root to leaves to create a matching graph, we then traverse it from leaves to root to determine the optimal solution for each root node assignment. To evaluate the matching score of a grounding, we use the maximum a posteriori probability described in \eqref{eqn.MAP_grounding}, where the score is the product of the distributions $P(o|\gamma_o)$ and $P(\gamma_o,\gamma_{o'}|o,r,o')$. For each leaf node in $T^*$, we merge nodes in $H$ with their parents, keeping the one which has the best score. We repeat this process until only mappings to root nodes are left, and then sort these root nodes by the score of the best tree which uses them.  This process is described in Algorithm \ref{alg:DP}, and has complexity of $O(|E^h|)$.

\begin{algorithm}[ht!]
  \caption{Solve for Optimal Groundings}
  \label{alg:DP}
  \begin{algorithmic}[1]
    \Procedure{Optimize groundings}{$T^*, H$}
    	\State{$Score(o)\triangleq P(o[0]|\gamma_o=o[1])$}
    	\State{$Score(o_1,o_2)\triangleq P(\gamma_{o_1}=o_1[1],\gamma_{o_2}=o_2[1]|o_1[0],r,o_2[0])$}
    	\State{Iterate from leaves to root}
      \ForAll{$o^* \in T^*$}
      	\If{$Parent(o^*)\neq \emptyset$}
      		\ForAll{$o \in O^h$ where $o[0]==Parent(o^*)$}
      			\State{$Score(o)*=\displaystyle max_{ c \in Children(o)}(Score(c)*Score(o,c))1(c[0]==o^*)$}
      		\EndFor
      	\EndIf
			\EndFor
    \EndProcedure
  \end{algorithmic}
\end{algorithm}

This process yields a set of groundings, $\gamma_{T^*}$, for each potential activity - generally on the order of the number of true groundings in the data. We then iterate through each grounding $\gamma \in \gamma_{T^*}$ and filtering groundings that have poor scores for the edges not present in the HPST $T^*$. In this way, we attempt to recover the grounding results for the original problem in \eqref{eqn.MAP_grounding} from the approximated problem \eqref{eqn.subgraph_thresh_grounding}. 
This allows us to have the speed of the HPSM approach and the effective quality of the full graph grounding results.

\section{Implementation} \label{sec:search_implementation}

In this section, we present implementation details of our approach. Fig.~\ref{fig:overview} shows an overview of our system. At a high-level, it operates as follows: as an archive video is recorded, detectors are applied to extract bounding boxes of objects of interest. These bounding boxes are then fused through a tracker and classified, yielding tracklets of objects that are stored in a table along with some simple attributes. During query-time, an analyst provides an ADSA query by an activity graph. From this query, an HPST is found according to \eqref{eqn.mdst}. The set of groundings that maximize \eqref{eqn.subgraph_thresh_grounding} are then found in the table as in Sec. \ref{sec:search_algorithm}. These groundings are scored and returned according to \eqref{eqn.grounding_model}.

\noindent \textbf{ADSA Query Vocabulary} 

We construct a vocabulary that corresponds to nodes and edges in the ADSA activity graph to allow for semantic descriptions of queries. We consider three classes of items, \textit{person}, \textit{object}, and \textit{vehicle}. Each item has a set of attributes that can be included in the query such as \textit{size}, \textit{appearing}, \textit{disappearing}, and \textit{speed}. Between each of these items, we define the following relationship attributes: \textit{same entity}, \textit{near}, \textit{not near}, and \textit{later}. The set of items and attributes can be expanded to include additional or more specific classes/descriptors. Due to the limited variety of objects in the datasets, we limit ourselves to simple semantic descriptors to prevent dominance of attributes in returns. 
By limiting the descriptiveness of attributes in our queries, we demonstrate retrieval capability in the presence of possible confusers. For a query such as ``two people loading an object into a pink truck'' a method that leverages primarily the color is not sufficiently general to handle ADSA's that do not include strong attribute descriptions. 

\noindent \textbf{Detection and Tracking}

We demonstrate the proposed method on three datasets. For the high quality VIRAT ground dataset \cite{oh2011large}, we use {P}iotr's {C}omputer {V}ision {M}atlab {T}oolbox \cite{PMT} to extract detections and then fuse them into tracklets \cite{andriyenko2012discrete}. In the case of the low-resolution WAMI AFRL data \cite{Castanon_mm15}, we apply algorithms designed specifically for aerial data \cite{xiao2010vehicle,wu2012multiple,divakaran2014real}. 
For the more complex AvA dataset \cite{gu2017ava}, a ResNet \cite{he2016deep} back-boned Faster RCNN \cite{ren2015faster} model was used for detection \cite{gu2017ava}, and the DeepSORT \cite{Wojke2017simple,Bewley2016_sort} algorithm was used for tracking.

\noindent \textbf{Relationship Learning} 

We learn semantic relationships by training a classifier on annotated positive and negative relationship examples of object pairs. For example, the relationship descriptor ``near'' between two items is found by training a classifier on features of two objects such as size, aspect ratios, distance between objects, etc. on a set of annotated examples of items that are near and are not near. Linear SVMs \cite{REF08a} are used to learn the relationships and provide the probabilities. 

\noindent \textbf{Re-ID}

Many of our queries requires maintaining identity over long periods of time, while tracked data inevitably has lost-tracks. We thus leverage re-identification ({\reid}) algorithms by utilizing a linear classifier ($f(X_1,X_2)=trace(WX_1 X_2^T)$) over the outer product of features $(X_1,X_2)$ from a pair of tracklets and train SVMs to learn $W$. This classifier is universally applied independent of context, pose, illumination etc. 

In practice, we have extremely limited training data that are properly annotated for the complex {\reid} models~\cite{zhang2014novel_reid,Das2014,Wang2016Person,Zheng2016Towards,Chen2018PAMI_REID}, 
so we use elementary target features like bounding box aspect ratios, locations, size, etc. 

\section{Experiments}
\label{sec:exp}

We perform semantic video retrieval experiments on three datasets: the VIRAT Ground 2.0 dataset \cite{oh2011large}, the AFRL Benchmark WAMI Data \cite{Castanon_mm15} and the Atomic Visual Actions (AvA) dataset \cite{gu2017ava}.
Given a set of activity graph queries, each algorithm is asked to return a ranked list of groundings in the archive video based on their likelihood scores. Each grounding is then represented by the minimal bounding spatio-temporal volume of the involved bounding boxes.
For VIRAT dataset where ground truth is provided, standard Precision-Recall curves are produced by varying the scoring threshold. 

\noindent
{\it Exploratory Search:} Our goal is to demonstrate the effectiveness of our method on large datasets. Typically, while ground-truth low-level detections are available, composite activities are seldom annotated. For this reason, we utilize a human to evaluate the returns and tabulate precision of top-k returns. We evaluate AFRL and AvA datasets in this context: a human operator evaluates the precision of top-k returns by watching the corresponding spatio-temporal window of the video. Each return is marked as a true detection if the overlap of the returned spatio-temporal volume and the true spatio-temporal volume is larger than 50\% of the union. 

As stated in Sec.~\ref{ssec:related_work}, most of the related methods \cite{tang2012learning,yang2013related,ma2014knowledge,Lin_2014_CVPR,Shu_2015_CVPR,won14,Choe_2013_ICCV} are not applicable to our setup as they retrieve relevant videos from a collection of short video snippets. 
We compare our performance with two approaches, a bag-of-words (BoW) scheme and a Manually Specified Graph Matching (MSGM) scheme. BoW is based on \cite{Lin_2014_CVPR,Shu_2015_CVPR}, it collects objects, object attributes and relationships in to a bag and ignores the structural relationships. 
To identify groundings, a bipartite matching scheme is utilized to find an assignment between the bag of words and a video snippet. We use our trained models for node-level concepts in this context.
For the MSGM method \cite{Castanon_mm15}, we quantify relationships by manually annotating data using bounding boxes for objects and then utilize subgraph matching of \cite{Castanon_mm15} on test data.

\subsection{Baseline Performance} \label{ssec:baseline}
We first show the baseline performance of three methods on human-annotated data of the VIRAT Ground 2.0 dataset \cite{oh2011large} with a set of seven queries. The VIRAT dataset is composed of 40 gigabytes of surveillance videos, capturing 11 scenes of  moving people and vehicles interacting. Resolution varies, with about 50$\times$100 pixels representing a pedestrian, and around 200$\times$200 pixels for vehicles.

\begin{figure*}[t]
\centering
\captionsetup[subfigure]{justification=centering}
    \begin{subfigure}[b]{0.32\textwidth}
      \centering  \includegraphics[width=.96\textwidth]{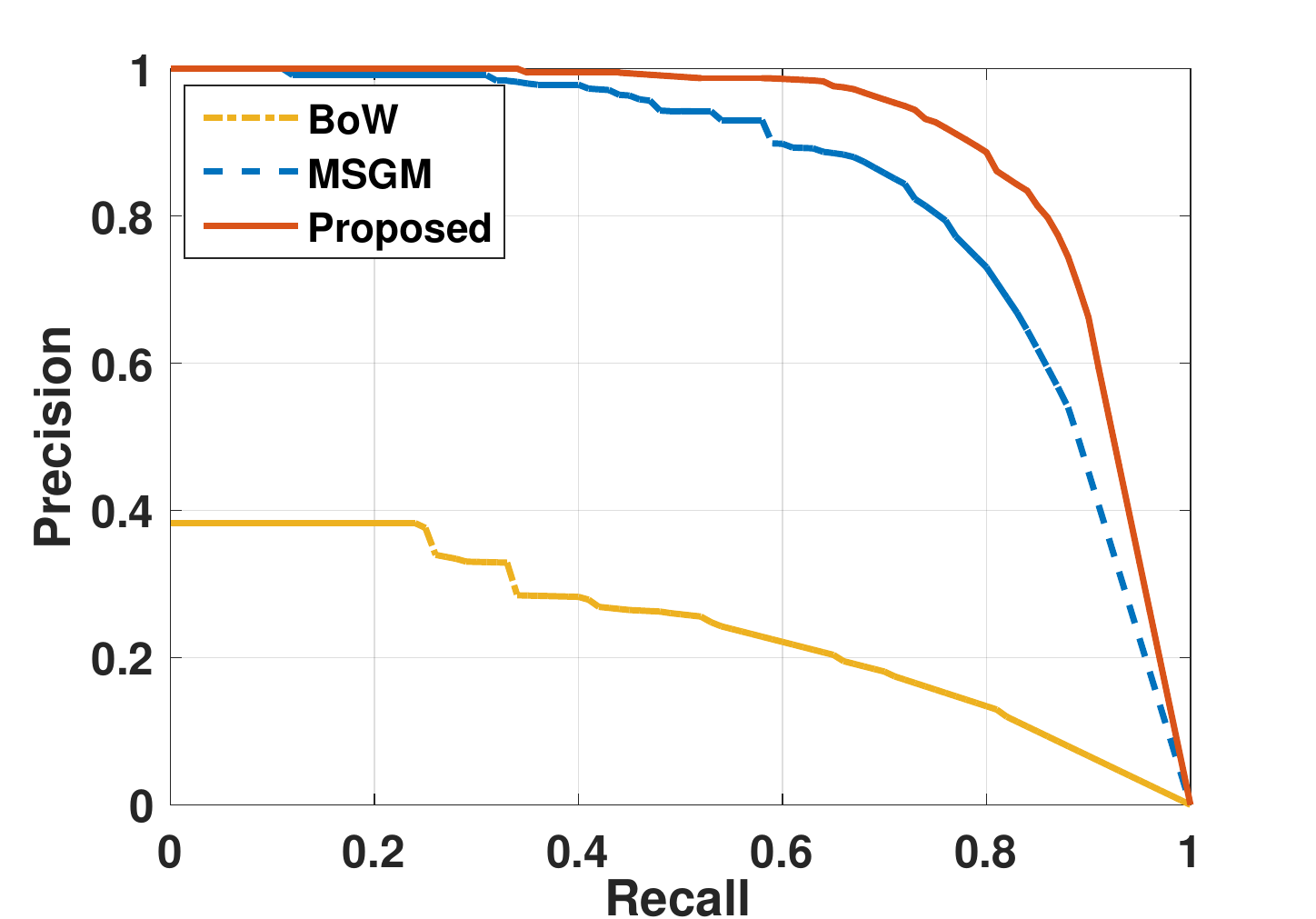}
        \caption{ Average ROC curve on VIRAT\\with human annotated data.}
        \label{fig:virat_roc_truth}
    \end{subfigure}
    \begin{subfigure}[b]{0.32\textwidth}
      \centering  \includegraphics[width=.96\textwidth]{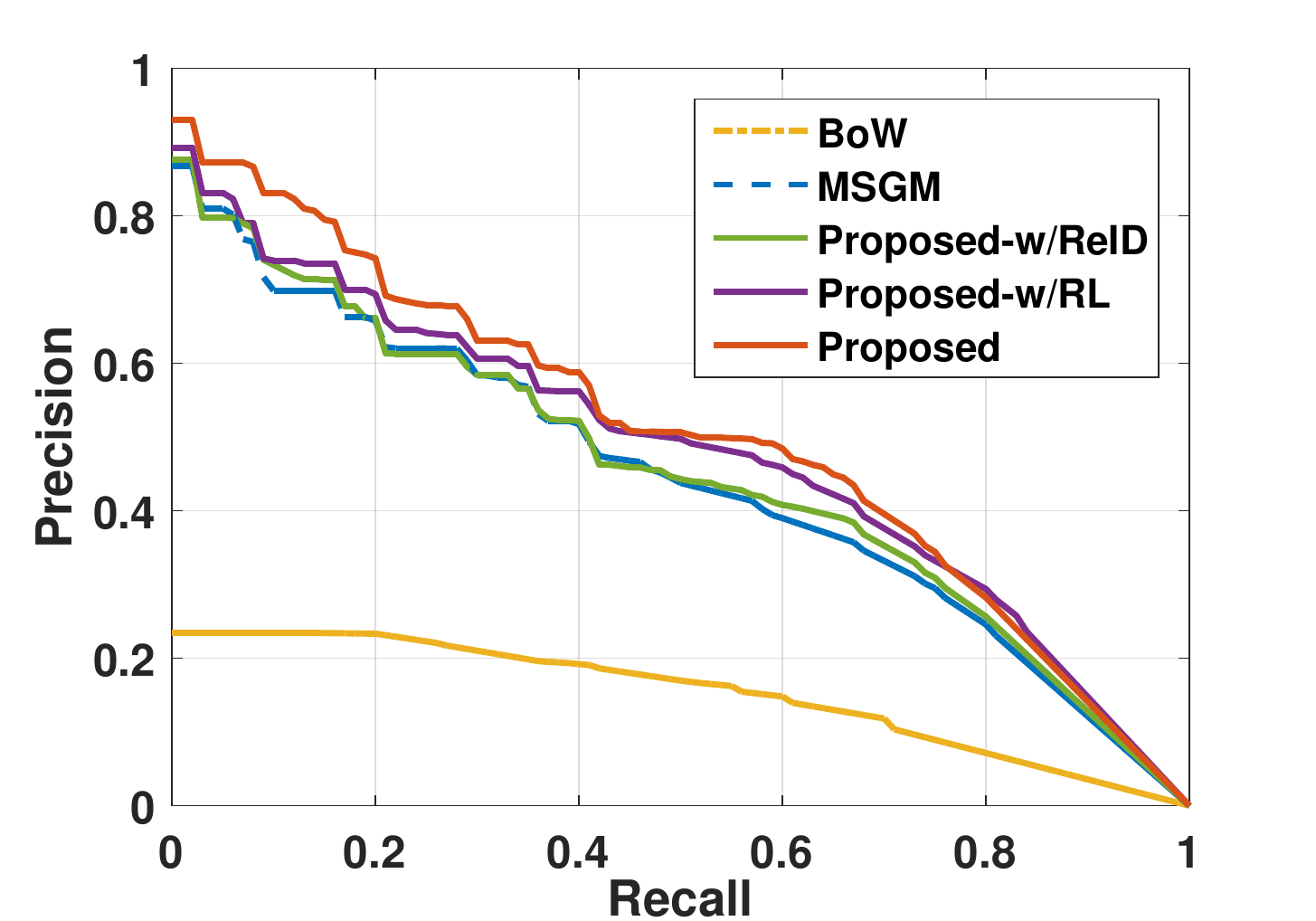}
        \caption{ Average ROC curve on VIRAT\\with tracked data.}
        \label{fig:virat_roc_track}
    \end{subfigure}
    \begin{subfigure}[b]{0.32\textwidth}
        \includegraphics[width=.96\textwidth]{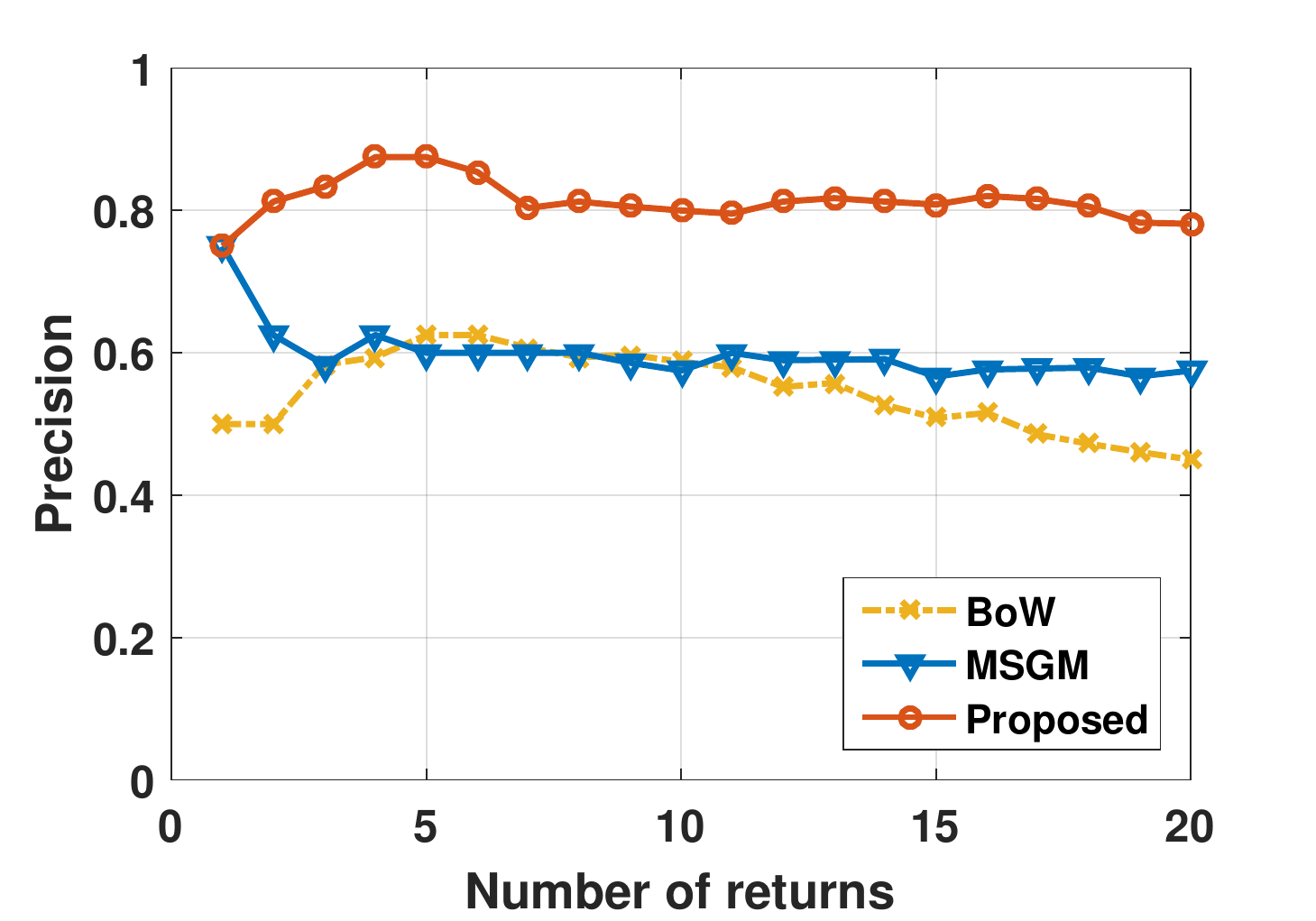}
       \centering
        \caption{ Average Precision wrt. number\\of returns on AFRL dataset.}
        \label{fig:yuma_prec}
    \end{subfigure}
    \caption{Retrieval performance}\label{fig:roc_curves}
\end{figure*}

As shown in Table \ref{tab:result-virat_GT} and Fig.~\ref{fig:virat_roc_truth}, the proposed approach outperforms BoW and MSGM.
On human annotated data, where we assume no uncertainty at the object level, we can see that both MSGM and the proposed method significantly outperform BoW. The queries all include some level of structural constraints between objects, for example, there is an underlying distance constraint for the people, car and object involved in {\it object deposit}. In a cluttered surveillance video where multiple activities occur at the same time, when an algorithm attempts to solve for a bipartite matching between people, car and objects, while ignoring the global spatial relationships between them, unrelated agents from different activities could be chosen, resulting in low detection accuracy for BoW. This shows that global structural relationships rather than isolated object-level descriptors are important. The performance gap between MGSM \cite{Castanon_mm15} and our method, indicates the importance of semantic concept learning and probabilistic reasoning 
over manually specified relationships and deterministic matching.

\begin{table}[t]
\renewcommand\arraystretch{1.2}
  \centering
  \begin{tabular}{|l|c|c|c|}
   \hline
     Query &  BoW \cite{Lin_2014_CVPR} & MSGM \cite{Castanon_mm15} & Proposed \\ \hline
    Person dismount & 15.33 & 78.26 & \textbf{83.93} \\
     Person mount & 21.37 & 70.61 & \textbf{83.94} \\
     Object deposit  & 26.39 & 71.34 & \textbf{85.69} \\
     Object take-out & 8.00 & 72.70 & \textbf{80.07} \\
     2 person deposit & 14.43 &  65.09 & \textbf{74.16} \\
     2 person take-out & 19.31 &  80.00 & \textbf{90.00} \\
     Group Meeting & 25.20 & 82.35 & \textbf{88.24} \\
     \hline    
     Average & 18.58  & 74.34 &  \textbf{83.72}  \\ 
     \hline
  \end{tabular}
  \caption{Area-Under-Curve (AUC) of precision-recall curves on VIRAT dataset with human annotated bounding boxes for Bag-of-Words approach (BoW \cite{Lin_2014_CVPR}), Manually Specified Graph Matching (MSGM \cite{Castanon_mm15}), and our proposed approach.}
  \label{tab:result-virat_GT}
\end{table}

\begin{table}[t]
\renewcommand\arraystretch{1.2}
  \centering
  \begin{tabular}{|l|c|c|c|c|c|}
   \hline
        \multirow{2}{*}{Query} &  \multirow{2}{*}{BoW  \cite{Lin_2014_CVPR}} & \multirow{2}{*}{MSGM \cite{Castanon_mm15}} & \multicolumn{3}{c|}{Proposed} \\ \cline{4-6}
        & & & Re-ID & RL & Full\\
        \hline
    Person dismount & 6.27  & 22.51  & 21.69 & 25.98 & \textbf{30.51} \\
    Person mount & 1.38  & 20.98 & 23.12 & 29.41 & \textbf{35.98} \\
    Object deposit & 7.90 & 46.27 &  47.79 & 47.62 & \textbf{49.13} \\
    Object take-out & 16.80 & 34.92 & 35.32 & 41.98 & \textbf{42.12} \\
    2 person deposit & 3.38 & 46.11 & 49.44 & \textbf{50.83} & \textbf{50.83}\\
    2 person take-out & 15.27 & 48.03 & 48.03 & \textbf{49.28} & \textbf{49.28} \\
    Group Meeting & 23.53 & 30.80  & 39.51 & 30.80 & \textbf{47.64} \\ \hline 
    Average & 10.65   &  35.66 & 37.84 & 39.41  & \textbf{43.64} \\
    \hline
  \end{tabular}
  \caption{Area-Under-Curve (AUC) of precision-recall curves on VIRAT dataset with automatically detected and tracked data for BoW \cite{Lin_2014_CVPR}, MSGM \cite{Castanon_mm15}, and our proposed approach with only {\reid} (Re-ID), with only relationship learning (RL), and the full system (Full) with both {\reid} and relationship learning. }
  \label{tab:result-virat_tracked}
\end{table}

\subsection{Probabilistic Reasoning with Noisy Input Data}
We perform an ablative analysis of our approach with detected and tracked bounding boxes in Table \ref{tab:result-virat_tracked} and Fig.~\ref{fig:virat_roc_track}. To demonstrate the effect of {\reid} and relationship learning, we report performance with only {\reid}, with only relationship learning, and with both {\reid} and relationship learning. 

Performance of all three methods degrade on tracked data due to miss detections/classifications and track errors. While ours degrades significantly, we still out-perform existing methods\footnote{Significant performance degradation with track data has also been observed in the context of activity classification even when full annotated data is available~\cite{Shu_2015_CVPR}.} for training an a priori known set of activities. 
For BoW, performance loss is large for the first six queries, due to reasons explained in Sec.~\ref{ssec:baseline}. 
Note that with BoW the group meeting query does not suffer significant degradation since it is more node-dominant than other queries (i.e, bipartite matching identifies multiple people present at the same time, and is a 
strong indicator of a meeting, particularly, in the absence of other co-occurring confusers).

\noindent
\textbf{Clutter v.s. Visual Distortion} 

On human annotated bounding boxes, we achieve an average AUC of 83.72\%. It indicates that our method is performing well in cluttered video free of visual distortions. Our performance drop to 43.64\% on tracked data is directly due to visual distortions introduced by miss-detections, miss-classifications and loss of tracks. This suggests that while our method compensates for some of the visual distortions, it is still important to improve detection, classification and tracking techniques.

\begin{figure}[h]
\centering
   \begin{subfigure}[b]{0.225\textwidth}
    \centering
        \includegraphics[width=.95\textwidth]{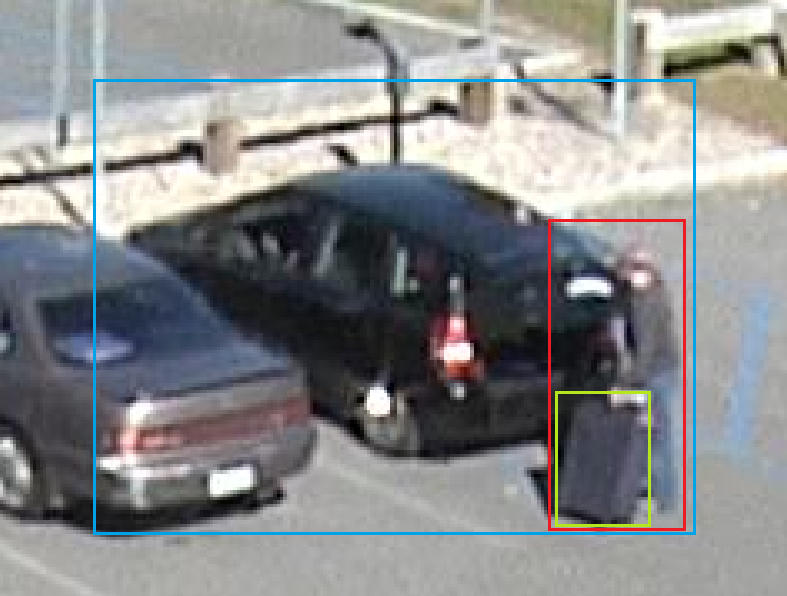}
        \caption{\small Detected obj. take-out}
        \label{fig:takeout_pos}
    \end{subfigure}
    \begin{subfigure}[b]{0.225\textwidth}
        \centering
        \includegraphics[width=.95\textwidth]{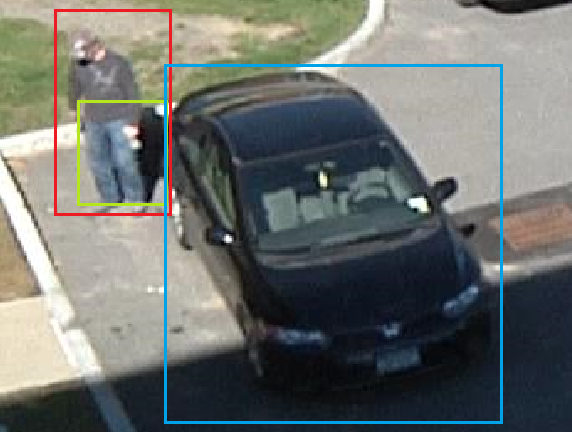}
        \caption{\small Rejected false obj. take-out}
        \label{fig:takeout_neg}
    \end{subfigure}
    \begin{subfigure}[b]{0.225\textwidth}   
    \centering
        \includegraphics[width=.95\textwidth]{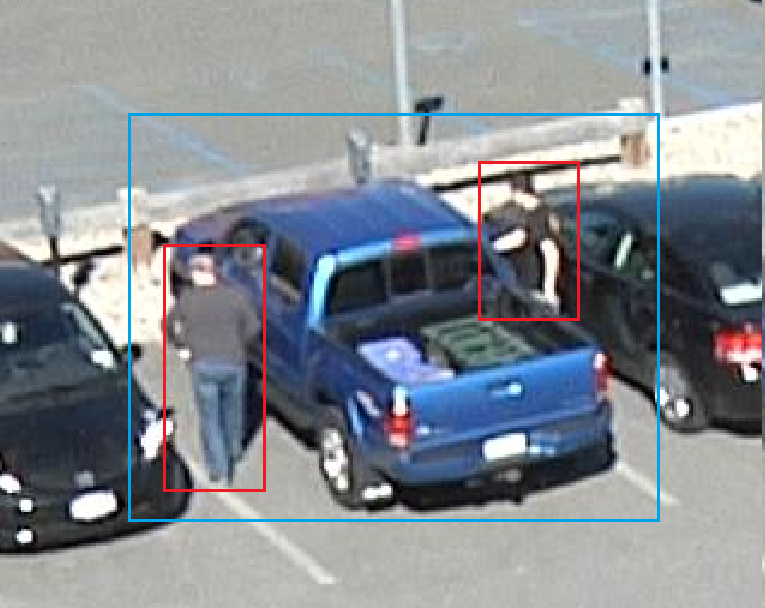}
        \caption{\small Detected mount return}
        \label{fig:mount_pos}
    \end{subfigure}
       \begin{subfigure}[b]{0.225\textwidth}
           \centering
        \includegraphics[width=.94\textwidth]{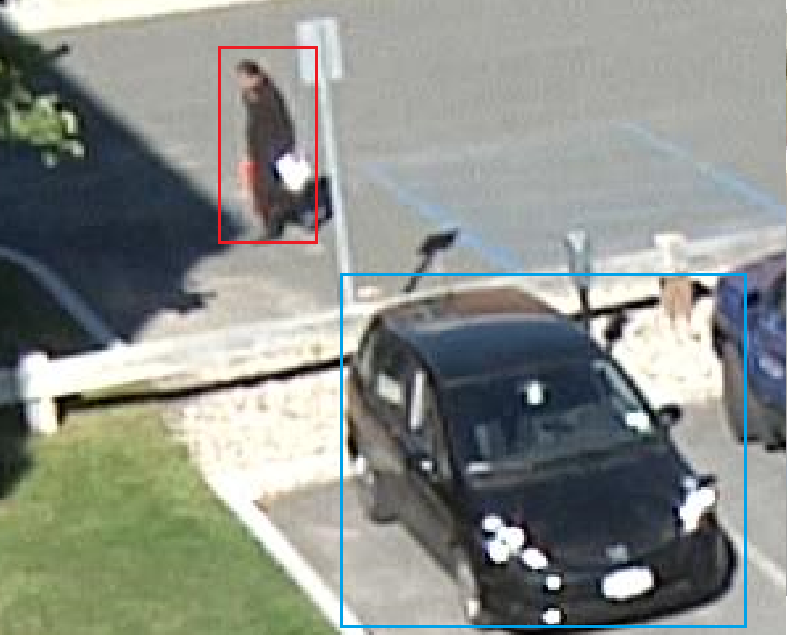}
        \caption{\small Rejected false mount return}
        \label{fig:mount_neg}
    \end{subfigure}
    \begin{subfigure}[b]{0.265\textwidth}
        \centering
    \includegraphics[width = .95\textwidth]{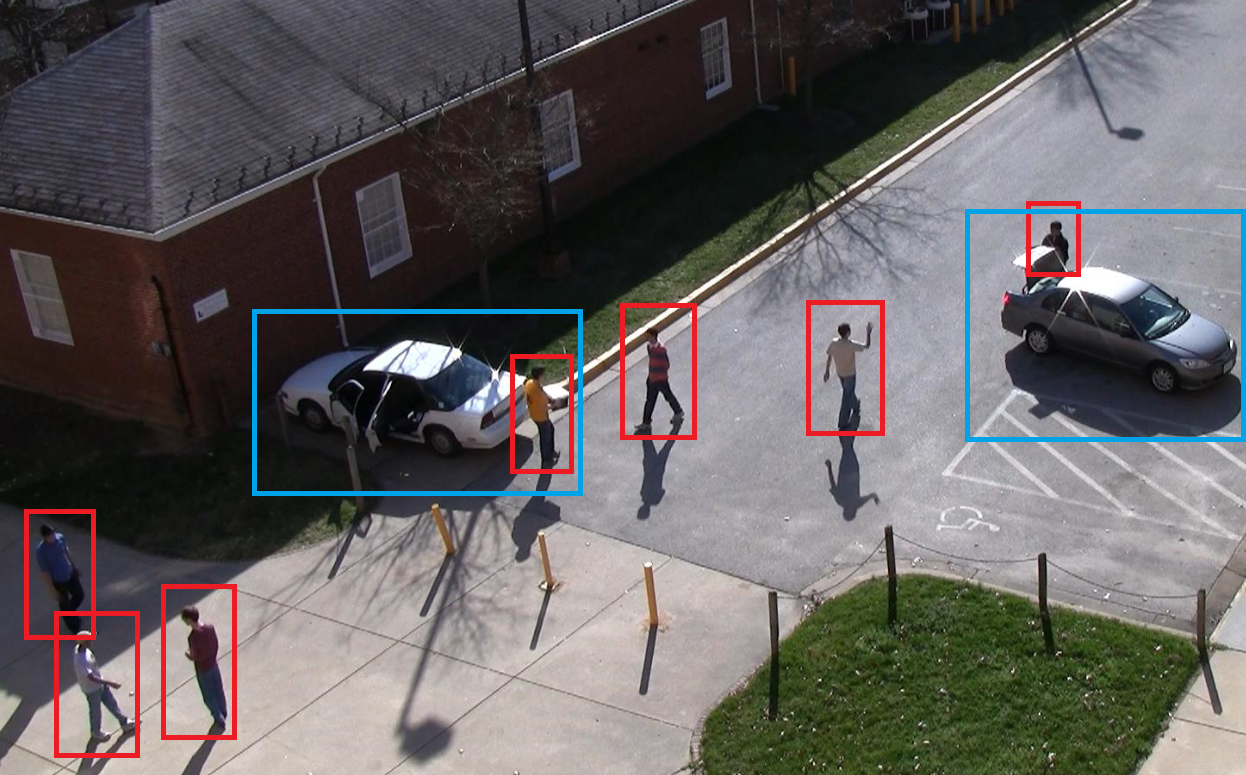}
    \caption{\small Clutter (VIRAT)}
            \label{fig:virat_clutter}
    \end{subfigure}
	\begin{subfigure}[b]{0.184\textwidth}
        \centering
    \includegraphics[width = .95\textwidth]{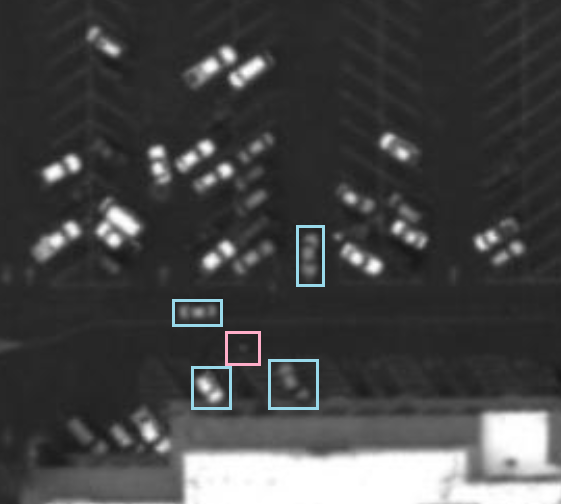}
    \caption{\small Clutter (AFRL)}
    \label{fig:yuma_clutter}
    \end{subfigure} 
    \caption{Retrieval Examples:  (b) and (d) show example of activities falsely returned by MSGM as \textit{obj. take-out} and \textit{person mount}. These activities are correctly rejected by our proposed approach. In (b), re-ID correctly stitches the suitcase tracks from before and after occlusion by the vehicle, and therefore does not return this as an example of \textit{obj. take-out}. In (d), MSGM describes the person as \textit{near} the vehicle, whereas our proposed approach does not, and therefore our approach does not return this as an example of \textit{person mount}. (e) and (f) demonstrate clutter present in the data, necessitating a retrieval system capable of reasoning over objects, attributes, and relationships.}\label{fig:examples}
\end{figure}

To visualize the importance of re-ID and relationship learning, we show examples of falsely returned MSGM outputs in Fig.~\ref{fig:examples}. In Figs.~\ref{fig:takeout_pos} and \ref{fig:mount_pos}, the objects are detected and tracked and both MSGM and our approach yield correct returns. In Fig.~\ref{fig:takeout_neg}, the suitcase is temporarily occluded by the vehicle. MSGM returns this as an example of \textit{object take-out}, as the suitcase is falsely described as \textit{appearing} after being occluded by the vehicle. Our proposed approach incorporates re-ID to classify the suitcase as the \textit{same} suitcase as prior to the occlusion, and therefore the suitcase is not described as \textit{appearing} and the example is rejected. Similarly, Fig.~\ref{fig:mount_neg} shows an MSGM false return for \textit{person mount} where a pedestrian walks by a car before the associated track is broken due to shadows. A manually input deterministic distance for \textit{near} across all perspectives leads to returning this as an example of \textit{person mount}. In contrast, our approach that learns an adaptive definition of \textit{near} identifies this relationship as \textit{not near} and correctly rejects this as an example of \textit{person mount}.

\subsection{Exploratory Search on WAMI Benchmark}

\begin{table*}[h]
\setlength{\tabcolsep}{4pt}
\renewcommand{\arraystretch}{1.2}
\centering
\begin{tabular}{|l|c|c|c|c|c|c|c|c|c|}
\hline
\multirow{2}{*}{Query} & \multicolumn{3}{|c|}{BoW \cite{Lin_2014_CVPR}} & \multicolumn{3}{|c|}{MSGM \cite{Castanon_mm15}} & \multicolumn{3}{|c|}{Proposed}  \\ \cline{2-10}

  & P@5 & P@10 & P@20 &P@5 & P@10 & P@20 & P@5 & P@10 & P@20 \\ \hline
  Car starts & 0.80 & 0.80 & 0.75 & 0.40 & 0.30 & 0.45 & 0.80 & 0.80 & 0.75\\
  Person mount & 0.60 & 0.80 & 0.50 & 0.80 & 0.80 & 0.60 & 1.00 & 0.80 & 0.75\\
  Car stops & 0.60 & 0.70 & 0.60 & 0.40 & 0.40 & 0.40 & 0.60 & 0.70 & 0.60\\
  Person dismount & 0.60 & 0.60 & 0.50 & 0.60 & 0.60 & 0.55 & 0.60 & 0.60 & 0.70\\
  \begin{tabular}[x]{@{}r@{}}Car suspicious\\stop\end{tabular} & 0.60 & 0.40 & 0.20 & 0.60 & 0.60 & 0.65 & 1.00 & 1.00 & 0.90  \\
  Car following & 0.40 & 0.40 & 0.30  & 0.60 & 0.50 & 0.55 & 0.80 & 0.70 & 0.70 \\
  \begin{tabular}[x]{@{}r@{}}Car following\\+stop\end{tabular}  & 0.60 & 0.40 & 0.40 & 0.80 & 0.50 & 0.60 & 1.00 & 0.70 & 0.80 \\
  \begin{tabular}[x]{@{}r@{}}Car following\\+dismount\end{tabular} & 0.60 & 0.50 & 0.30 & 0.80 & - & - &1.00 &1.00 & - \\  
\hline
\end{tabular}
  \caption{Precision @ top-k return results for AFRL aerial benchmark dataset.}
  \label{tab:YUMA_precision}
\end{table*}

The AFRL Benchmark WAMI data is from a wide-area persistent surveillance sensor flying over  $\approx$4 sq. km in Yuma, AZ. It contains 110 minutes of large (8000 $\times$ 8000), low-contrast, low frame rate (1.5 fps), low resolution (0.25 m/pixel) gray scale imagery. Vehicles and people roughly occupy approximately 50-150  and 10 pixels, respectively, leading to noisy detector/tracker outputs.

We search for queries of varying complexity. Simple queries like \textit{car starts} and \textit{car stops} where a stationary car starts moving or a moving vehicle comes to a prolonged stop, can be described by a single node with corresponding attributes. \textit{Person mount} and \textit{person dismount} are built on top of the single node car queries by adding a person getting into or out of an vehicle. Complex queries like \textit{car suspicious stop} searches for a car that comes to a stop for a period of time then continues moving. 
Finally, we search for composite queries, \textit{car following + stop}, a \textit{car following} activity immediately proceeded by a \textit{car suspicious stop} activity, and \textit{car following + dismount}, a \textit{car following} activity immediately proceeded by a \textit{person dismount} activity.

We compare performance of our proposed approach to BoW and MSGM in Table \ref{tab:YUMA_precision} and Fig.~\ref{fig:yuma_prec}. Ground truth labeling is unavailable for this dataset, so we report performance as precision at $k$ for $k=5,10, \mbox{ and } 20$.

\noindent
{\textbf{Dominant vs. Weak Attributes} 

We can see that BoW outperforms MSGM for simple queries like {\it car starts} and {\it car stops} where a dominant signature of object attribute is present. It is reasonable since BoW learns attribute classifiers for car starting or stopping based on the speed of the vehicle, while MSGM uses a manually specified speed constraint. In contrast, when multiple agents are involved and thus the structural relationships between agents compose part of the query, MSGM outperforms BoW. It suggests the need for reasoning with relationships between objects to capture the activity. The proposed approach combines the attribute learning from BoW, along with additional ability to learn semantic relationships, and as such, outperforms both BoW and MSGM. Our performance gain is more significant on complex composite queries like {\it car following + stop} or {\it car following + dismount}, which demonstrates the benefits from different components of our system.}

\noindent
\textbf{Co-occurring Activities}

Figs.~\ref{fig:virat_clutter} and \ref{fig:yuma_clutter} demonstrate the importance of grounding when many other unrelated co-occurring activities are present in the data, which leads to significant degradation with BoW based approaches. For these scenarios, a retrieval system must be able to reason with objects, attributes and relationships to find the correct grounding that matches the query.

\subsection{Exploratory Search on Real-life Movie Videos}

To demonstrate the applicability of our method to various data sources, we perform exploratory search on a recently published video dataset of Atomic Visual Actions (AVA) \cite{gu2017ava}. The AVA dataset consists of 430 15-minute video clips from movies for action recognition of 80 atomic action classes like stand, sit, talk to, dance, ride, eat, and so on. This dataset has multi-person activities occurring in realistic scenes. 

Nevertheless, like the authors point out, the focus of AVA dataset is to detect/recognize atomic visual actions that manifest over a short period of time. On the other hand, our focus is on retrieving sparsely or unannotated composite activities composed of low-level atomic actions. Yet, we seek to highlight the utility of our scheme on real-life videos, although, our scheme is focused on identification of long-term events with multi-agent activities and assumes availability of low-level classifiers, detectors and trackers.

We construct composite queries based on atomic actions present in the dataset. \textit{Hug then kiss} corresponds to an activity where two people hug each other for some time, then kiss each other. \textit{Group meeting} involves three or more people talking or listening to each other while sitting or standing for some period of time. \textit{Dancing party} refers to multiple people hugging and dancing in the same scene. As for \textit{band performance}, multiple people are singing, dancing and playing musical instruments. These composite activities are composed of atomic actions such as talk, listen, stand, dance, hug, sing and so on. Note that a subset of atomic activities are shared among the composite activities.

We evaluate the algorithms on the 64 validation video clips, both ground truth atomic action labels, as well as action detection results are used as input. As no identity label is provided, the same tracking algorithm is applied to both input data to associate actions between frames. Since ground truth for the composite queries is not available, we manually evaluate the performance for precision at $k$ for $k$= 1,5, and 10. We observe that for queries other than \textit{group meeting} (there are lots of group meetings in these videos), the average number of returns are from 5 to 15 returns. Note that the results are usually round numbers as the maximum number of returns we look at is 10.

\begin{table*}
\setlength{\tabcolsep}{4pt}
\renewcommand{\arraystretch}{1.2}
\centering
\begin{tabular}{|l|l|c|c|c|c|c|c|c|c|c|}
\hline
 & \multirow{2}{*}{Query} & \multicolumn{3}{|c|}{BoW \cite{Lin_2014_CVPR}} & \multicolumn{3}{|c|}{MSGM \cite{Castanon_mm15}} & \multicolumn{3}{|c|}{Proposed}  \\ \cline{3-11}

  & & P@1 & P@5 & P@10 & P@1 & P@5 & P@10 & P@1 & P@5 & P@10
  \\ \hline
\multirow{4}{*}{\begin{tabular}[x]{@{}r@{}}Annotated\\atomic\\actions\end{tabular}}  &  Hug then kiss & 1.00 & 1.00 & 0.80 & 1.00 & 1.00 & 1.00 & 1.00 & 1.00 & 1.00\\
 & Group meeting & 1.00 & 0.60 & 0.60 & 1.00 & 0.80 & 0.80 & 1.00 & 0.80 & 0.90  \\
 & Dancing party & 1.00 & 0.80 & -  & 1.00 & 1.00 & - & 1.00 & 1.00 & - \\
 & Band performance  & 1.00 & 0.80 & 0.80 & 1.00 & 0.80 & 0.90 & 1.00 & 0.80 & 0.90 \\ 
\hline
\hline
\multirow{4}{*}{\begin{tabular}[x]{@{}r@{}}Detected\\atomic\\actions\end{tabular}}  &  Hug then kiss & 1.00 & 0.80 & 0.60 & 1.00 & 0.80 & - & 1.00 & 0.80 & -\\
 & Group meeting & 1.00 & 0.60 & 0.40 & 1.00 & 0.80 & 0.60 & 1.00 & 0.80 & 0.70  \\
 & Dancing party & 0.00 & 0.00 & -  & 0.00 & 0.33 & - & 0.00 & 0.33 & - \\
 & Band performance  & 0.00 & 0.40 & - & 1.00 & 0.40 & - & 1.00 & 0.40 & - \\  
\hline
\end{tabular}
  \caption{Precision @ top-k return results for AvA benchmark dataset. "Annotated atomic actions" means human annotated atomic actions are used as input. "Detected atomic actions" means atomic actions detected by an action detector \cite{gu2017ava} are used as input. Both inputs are tracked by DeepSORT \cite{Wojke2017simple,Bewley2016_sort} to generate the identity associations.}
  \label{tab:AvA_precision}
\end{table*}

\noindent
\textbf{Structural Relationships}

The results on AVA are consistent with what we observe for VIRAT and AFRL datasets. In all cases the algorithms have similar performance for short term activities like \textit{hug then kiss}. Errors made by BoW on annotated actions are due to the fact that the temporal order of two atomic actions is not considered.

The proposed method and MSGM performs better in longer-term composite activities than the BoW method. In these cases the use of structural relationships between atomic actions is important in correctly recognizing the activity. The performance gap is due to the lack of identity relationships between actions in BoW.

We observe smaller performance gap between MSGM and the proposed method here, since, in movie-style videos the distance between people has smaller range (variability) and so a fixed preset distance threshold works reasonably well.  In contrast, in surveillance scenarios the variability is significantly larger leading to larger errors for MSGM in comparison to our method. 

\noindent
\textbf{Visual Distortions}

The comparisons between annotated action labels and detected action results, reveals that detectors work fairly well in detecting actions with strong visual signature like "kiss", while accuracy drops for actions like "sing", "talk" and "dance", which leads to the performance drops in the composite activities. In particular, we find that the precision for \textit{dancing party} is very low for all methods, as the action detector is unable to differentiate "dance" and "hug".
We also observe that fewer results were returned for \textit{dancing party} and \textit{dance performance} from the detected action data than the annotated data, indicating the loss of detected actions in the first place.

We observe some incorrect returns from annotated atomic actions in composite activities, and these returns are related to broken tracks. For \textit{group meeting} where at least three people with different identities are required, some of the returns only involved two people. Inaccurate detections along with broken tracks contribute to degradation in precision of retrieved activities.

\section{Summary}
In this paper, we incorporate component level similarity to the problem of semantic activity retrieval in large surveillance videos. We represent semantic queries by activity graphs and propose a novel probabilistic approach to efficiently identify potential spatio-temporal locations to ground activity graphs in cluttered videos. Our experiments show superior performance over methods that fail to consider structural relationships between objects or ignore input data noise and domain-specific variance. The proposed method is robust to visual distortions  and capable of suppressing clutter that is inevitable in surveillance videos.


%





\ifCLASSOPTIONcaptionsoff
  \newpage
\fi



\bibliographystyle{IEEEtran}
\bibliography{IEEEabrv,search}

\begin{thebibliography}{10}
\providecommand{\url}[1]{#1}
\csname url@samestyle\endcsname
\providecommand{\newblock}{\relax}
\providecommand{\bibinfo}[2]{#2}
\providecommand{\BIBentrySTDinterwordspacing}{\spaceskip=0pt\relax}
\providecommand{\BIBentryALTinterwordstretchfactor}{4}
\providecommand{\BIBentryALTinterwordspacing}{\spaceskip=\fontdimen2\font plus
\BIBentryALTinterwordstretchfactor\fontdimen3\font minus
  \fontdimen4\font\relax}
\providecommand{\BIBforeignlanguage}[2]{{%
\expandafter\ifx\csname l@#1\endcsname\relax
\typeout{** WARNING: IEEEtran.bst: No hyphenation pattern has been}%
\typeout{** loaded for the language `#1'. Using the pattern for}%
\typeout{** the default language instead.}%
\else
\language=\csname l@#1\endcsname
\fi
#2}}
\providecommand{\BIBdecl}{\relax}
\BIBdecl

\bibitem{Yu2016retrieval}
L.~Yu, Z.~Huang, J.~Cao, and H.~T. Shen, ``Scalable video event retrieval by
  visual state binary embedding,'' \emph{IEEE Transactions on Multimedia},
  vol.~18, no.~8, pp. 1590--1603, Aug 2016.

\bibitem{Liong2017deepvideo}
V.~E. Liong, J.~Lu, Y.~P. Tan, and J.~Zhou, ``Deep video hashing,'' \emph{IEEE
  Transactions on Multimedia}, vol.~19, no.~6, pp. 1209--1219, June 2017.

\bibitem{Lin2017deepretrieval}
J.~Lin, L.~Y. Duan, S.~Wang, Y.~Bai, Y.~Lou, V.~Chandrasekhar, T.~Huang,
  A.~Kot, and W.~Gao, ``Hnip: Compact deep invariant representations for video
  matching, localization, and retrieval,'' \emph{IEEE Transactions on
  Multimedia}, vol.~19, no.~9, pp. 1968--1983, Sept 2017.

\bibitem{Johnson}
J.~Johnson, R.~Krishna, M.~Stark, L.-J. Li, D.~Shamma, M.~Bernstein, and
  L.~Fei-Fei, ``Image retrieval using scene graphs,'' in \emph{Proceedings of
  the IEEE Conference on Computer Vision and Pattern Recognition}, 2015, pp.
  3668--3678.

\bibitem{Castanon_mm15}
G.~Casta{\~n}{\'o}n, Y.~Chen, Z.~Zhang, and V.~Saligrama, ``Efficient activity
  retrieval through semantic graph queries,'' in \emph{Proceedings of the 23rd
  Annual ACM Conference on Multimedia Conference}.\hskip 1em plus 0.5em minus
  0.4em\relax ACM, 2015, pp. 391--400.

\bibitem{Lin_2014_CVPR}
D.~Lin, S.~Fidler, C.~Kong, and R.~Urtasun, ``Visual semantic search:
  Retrieving videos via complex textual queries,'' in \emph{Proceedings of the
  IEEE Conference on Computer Vision and Pattern Recognition}, June 2014, pp.
  2657--2664.

\bibitem{Xu_2015_CVPR}
Z.~Xu, Y.~Yang, and A.~G. Hauptmann., ``A discriminative cnn video
  representation for event detection.'' in \emph{Proceedings of the IEEE
  Conference on Computer Vision and Pattern Recognition}, 2015, pp. 1798--1807.

\bibitem{Jain_2015_CVPR}
M.~Jain, J.~van Gemert, and C.~Snoek., ``What do 15,000 object categories tell
  us about classifying and localizing actions?'' in \emph{Proceedings of the
  IEEE Conference on Computer Vision and Pattern Recognition}, 2015, pp.
  46--55.

\bibitem{karpathy_2014_CVPR}
A.~Karpathy, G.~Toderici, S.~Shetty, T.~Leung, R.~Sukthankar, and L.~Fei-Fei.,
  ``Large-scale video classification with convolutional neural networks.'' in
  \emph{Proceedings of the IEEE Conference on Computer Vision and Pattern
  Recognition}, 2014, pp. 1725--1732.

\bibitem{Simonyan_NIPS_2014}
K.~Simonyan and A.~Zisserman., ``Two-stream convolutional networks for action
  recognition in videos.'' in \emph{Advances in Neural Information Processing
  Systems}, 2014.

\bibitem{tang2012learning}
K.~Tang, L.~Fei-Fei, and D.~Koller, ``Learning latent temporal structure for
  complex event detection,'' in \emph{Proceedings of the IEEE Conference on
  Computer Vision and Pattern Recognition}.\hskip 1em plus 0.5em minus
  0.4em\relax IEEE, 2012, pp. 1250--1257.

\bibitem{yang2013related}
Y.~Yang, Z.~Ma, Z.~Xu, S.~Yan, and A.~G. Hauptmann, ``How related exemplars
  help complex event detection in web videos?'' in \emph{Proceedings of the
  IEEE International Conference on Computer Vision}, 2013, pp. 2104--2111.

\bibitem{ma2014knowledge}
Z.~Ma, Y.~Yang, N.~Sebe, and A.~G. Hauptmann, ``Knowledge adaptation with
  partially shared features for event detection using few exemplars,''
  \emph{IEEE Transactions on Pattern Analysis and Machine Intelligence},
  vol.~36, no.~9, pp. 1789--1802, 2014.

\bibitem{Shu_2015_CVPR}
T.~Shu, D.~Xie, B.~Rothrock, S.~Todorovic, and S.~Chun~Zhu, ``Joint inference
  of groups, events and human roles in aerial videos,'' in \emph{Proceedings of
  the IEEE Conference on Computer Vision and Pattern Recognition}, 2015, pp.
  4576--4584.

\bibitem{won14}
W.~Choi and S.~Savarese, ``Understanding collective activities of people from
  videos,'' \emph{IEEE Transactions on Pattern Analysis and Machine
  Intelligence}, vol.~36, no.~6, pp. 1242--1257, 2014.

\bibitem{wang2014event}
F.~Wang, Z.~Sun, Y.~G. Jiang, and C.~W. Ngo, ``Video event detection using
  motion relativity and feature selection,'' \emph{IEEE Transactions on
  Multimedia}, vol.~16, no.~5, pp. 1303--1315, Aug 2014.

\bibitem{Ye2015deepvideo}
H.~Ye, Z.~Wu, R.-W. Zhao, X.~Wang, Y.-G. Jiang, and X.~Xue, ``Evaluating
  two-stream cnn for video classification,'' in \emph{Proceedings of the 5th
  ACM on International Conference on Multimedia Retrieval}, ser. ICMR
  '15.\hskip 1em plus 0.5em minus 0.4em\relax New York, NY, USA: ACM, 2015, pp.
  435--442.

\bibitem{zhao2017videowhisper}
N.~Zhao, H.~Zhang, R.~Hong, M.~Wang, and T.-S. Chua, ``Videowhisper: Toward
  discriminative unsupervised video feature learning with attention-based
  recurrent neural networks,'' \emph{IEEE Transactions on Multimedia}, vol.~19,
  no.~9, pp. 2080--2092, 2017.

\bibitem{zhang2018discriminative}
S.~Zhang, C.~Gao, J.~Zhang, F.~Chen, and N.~Sang, ``Discriminative part
  selection for human action recognition,'' \emph{IEEE Transactions on
  Multimedia}, vol.~20, no.~4, pp. 769--780, 2018.

\bibitem{wang2018two}
X.~Wang, L.~Gao, P.~Wang, X.~Sun, and X.~Liu, ``Two-stream 3-d convnet fusion
  for action recognition in videos with arbitrary size and length,'' \emph{IEEE
  Transactions on Multimedia}, vol.~20, no.~3, pp. 634--644, 2018.

\bibitem{kumar2018f}
K.~Kumar and D.~D. Shrimankar, ``F-des: Fast and deep event summarization,''
  \emph{IEEE Transactions on Multimedia}, vol.~20, no.~2, pp. 323--334, 2018.

\bibitem{Hoi2008Multi}
S.~C.~H. Hoi and M.~R. Lyu, ``A multimodal and multilevel ranking scheme for
  large-scale video retrieval,'' \emph{IEEE Transactions on Multimedia},
  vol.~10, no.~4, pp. 607--619, June 2008.

\bibitem{Xu2008Multi}
C.~Xu, Y.~F. Zhang, G.~Zhu, Y.~Rui, H.~Lu, and Q.~Huang, ``Using webcast text
  for semantic event detection in broadcast sports video,'' \emph{IEEE
  Transactions on Multimedia}, vol.~10, no.~7, pp. 1342--1355, Nov 2008.

\bibitem{Chen2012Multi}
X.~Chen, A.~O.~H. III, and S.~Savarese, ``Multimodal video indexing and
  retrieval using directed information,'' \emph{IEEE Transactions on
  Multimedia}, vol.~14, no.~1, pp. 3--16, Feb 2012.

\bibitem{Merler2012event}
M.~Merler, B.~Huang, L.~Xie, G.~Hua, and A.~Natsev, ``Semantic model vectors
  for complex video event recognition,'' \emph{IEEE Transactions on
  Multimedia}, vol.~14, no.~1, pp. 88--101, Feb 2012.

\bibitem{Ma2013event}
Z.~Ma, Y.~Yang, N.~Sebe, K.~Zheng, and A.~G. Hauptmann, ``Multimedia event
  detection using a classifier-specific intermediate representation,''
  \emph{IEEE Transactions on Multimedia}, vol.~15, no.~7, pp. 1628--1637, Nov
  2013.

\bibitem{Mazloom2016event}
M.~Mazloom, X.~Li, and C.~G.~M. Snoek, ``Tagbook: A semantic video
  representation without supervision for event detection,'' \emph{IEEE
  Transactions on Multimedia}, vol.~18, no.~7, pp. 1378--1388, July 2016.

\bibitem{Song2017event}
H.~Song, X.~Wu, W.~Yu, and Y.~Jia, ``Extracting key segments of videos for
  event detection by learning from web sources,'' \emph{IEEE Transactions on
  Multimedia}, vol.~PP, no.~99, pp. 1--1, 2017.

\bibitem{Pang2015retrieval}
L.~Pang, S.~Zhu, and C.~W. Ngo, ``Deep multimodal learning for affective
  analysis and retrieval,'' \emph{IEEE Transactions on Multimedia}, vol.~17,
  no.~11, pp. 2008--2020, Nov 2015.

\bibitem{shi2017sequential}
Y.~Shi, Y.~Tian, Y.~Wang, and T.~Huang, ``Sequential deep trajectory descriptor
  for action recognition with three-stream cnn,'' \emph{IEEE Transactions on
  Multimedia}, vol.~19, no.~7, pp. 1510--1520, 2017.

\bibitem{Hao2017hashing}
Y.~Hao, T.~Mu, J.~Y. Goulermas, J.~Jiang, R.~Hong, and M.~Wang, ``Unsupervised
  t-distributed video hashing and its deep hashing extension,'' \emph{IEEE
  Transactions on Image Processing}, vol.~26, no.~11, pp. 5531--5544, Nov 2017.

\bibitem{han2017vrfp}
X.~Han, B.~Singh, V.~I. Morariu, and L.~S. Davis, ``Vrfp: On-the-fly video
  retrieval using web images and fast fisher vector products,'' \emph{IEEE
  Transactions on Multimedia}, vol.~19, no.~7, pp. 1583--1595, July 2017.

\bibitem{wu2014zero}
S.~Wu, S.~Bondugula, F.~Luisier, X.~Zhuang, and P.~Natarajan, ``Zero-shot event
  detection using multi-modal fusion of weakly supervised concepts,'' in
  \emph{Proceedings of the IEEE Conference on Computer Vision and Pattern
  Recognition}, 2014, pp. 2665--2672.

\bibitem{chang2015semantic}
X.~Chang, Y.~Yang, A.~G. Hauptmann, E.~P. Xing, and Y.-L. Yu, ``Semantic
  concept discovery for large-scale zero-shot event detection,'' in
  \emph{Proceedings of the 24th International Conference on Artificial
  Intelligence}, 2015, pp. 2234--2240.

\bibitem{elhoseiny2015zero}
M.~Elhoseiny, J.~Liu, H.~Cheng, H.~Sawhney, and A.~Elgammal, ``Zero-shot event
  detection by multimodal distributional semantic embedding of videos,'' in
  \emph{Proceedings of the 30th AAAI Conference on Artificial Intelligence},
  2016.

\bibitem{gan2015exploring}
C.~Gan, M.~Lin, Y.~Yang, Y.~Zhuang, and A.~G. Hauptmann, ``Exploring semantic
  inter-class relationships (sir) for zero-shot action recognition,'' in
  \emph{Proceedings of the 29th AAAI Conference on Artificial Intelligence},
  2015, pp. 3769--3775.

\bibitem{Jain_2015_ICCV}
M.~Jain, J.~C. van Gemert, T.~Mensink, and C.~G.~M. Snoek, ``Objects2action:
  Classifying and localizing actions without any video example,'' in
  \emph{Proceedings of the IEEE International Conference on Computer Vision},
  December 2015.

\bibitem{Zhang_2016_CVPR}
Y.~Zhang, B.~Gong, and M.~Shah, ``Fast zero-shot image tagging,'' in
  \emph{Proceedings of the IEEE Conference on Computer Vision and Pattern
  Recognition}, June 2016, pp. 5985--5994.

\bibitem{lampert2009attribute}
C.~H. Lampert, H.~Nickisch, and S.~Harmeling, ``Learning to detect unseen
  object classes by between-class attribute transfer,'' in \emph{Proceedings of
  the IEEE Conference on Computer Vision and Pattern Recognition}, 2009, pp.
  951--958.

\bibitem{wang2016zeroshot-reid}
Z.~Wang, R.~Hu, C.~Liang, Y.~Yu, J.~Jiang, M.~Ye, J.~Chen, and Q.~Leng,
  ``Zero-shot person re-identification via cross-view consistency,'' \emph{IEEE
  Transactions on Multimedia}, vol.~18, no.~2, pp. 260--272, Feb 2016.

\bibitem{Choe_2013_ICCV}
T.~E. Choe, H.~Deng, F.~Guo, M.~W. Lee, and N.~Haering., ``Semantic
  video-to-video search using sub-graph grouping and matching.'' in
  \emph{Proceedings of the IEEE International Conference on Computer Vision},
  2013.

\bibitem{zitnick}
D.~F. Fouhey and C.~L. Zitnick, ``Predicting object dynamics in scenes,'' in
  \emph{Proceedings of the IEEE Conference on Computer Vision and Pattern
  Recognition}, June 2014, pp. 2019--2026.

\bibitem{cheng2014bing}
M.-M. Cheng, Z.~Zhang, W.-Y. Lin, and P.~Torr, ``Bing: Binarized normed
  gradients for objectness estimation at 300fps,'' in \emph{Proceedings of the
  IEEE Conference on Computer Vision and Pattern Recognition}, 2014, pp.
  3286--3293.

\bibitem{Lafferty:2001:CRF:645530.655813}
J.~D. Lafferty, A.~McCallum, and F.~C.~N. Pereira, ``Conditional random fields:
  Probabilistic models for segmenting and labeling sequence data,'' in
  \emph{Proceedings of the 18th International Conference on Machine Learning},
  2001.

\bibitem{platt1999probabilistic}
J.~Platt \emph{et~al.}, ``Probabilistic outputs for support vector machines and
  comparisons to regularized likelihood methods,'' \emph{Advances in Large
  Margin Classifiers}, vol.~10, no.~3, pp. 61--74, 1999.

\bibitem{castanon2012exploratory}
G.~D. Castanon, A.~L. Caron, V.~Saligrama, and P.-m. Jodoin, ``Exploratory
  search of long surveillance videos,'' in \emph{Proceedings of the 20th Annual
  ACM International Conference on Multimedia}.\hskip 1em plus 0.5em minus
  0.4em\relax ACM, 2012, pp. 309--318.

\bibitem{oh2011large}
S.~Oh, A.~Hoogs, A.~Perera, N.~Cuntoor, C.-C. Chen, J.~T. Lee, S.~Mukherjee,
  J.~Aggarwal, H.~Lee, L.~Davis \emph{et~al.}, ``A large-scale benchmark
  dataset for event recognition in surveillance video,'' in \emph{Proceedings
  of the IEEE Conference on Computer Vision and Pattern Recognition}.\hskip 1em
  plus 0.5em minus 0.4em\relax IEEE, 2011, pp. 3153--3160.

\bibitem{PMT}
P.~Doll\'ar, ``{P}iotr's {C}omputer {V}ision {M}atlab {T}oolbox ({PMT}),''
  \url{https://github.com/pdollar/toolbox}.

\bibitem{andriyenko2012discrete}
A.~Andriyenko, K.~Schindler, and S.~Roth, ``Discrete-continuous optimization
  for multi-target tracking,'' in \emph{Proceedings of the IEEE Conference on
  Computer Vision and Pattern Recognition}.\hskip 1em plus 0.5em minus
  0.4em\relax IEEE, 2012, pp. 1926--1933.

\bibitem{xiao2010vehicle}
J.~Xiao, H.~Cheng, H.~Sawhney, and F.~Han, ``Vehicle detection and tracking in
  wide field-of-view aerial video,'' in \emph{Proceedings of the IEEE
  Conference on Computer Vision and Pattern Recognition}.\hskip 1em plus 0.5em
  minus 0.4em\relax IEEE, 2010, pp. 679--684.

\bibitem{wu2012multiple}
S.~Wu, S.~Das, Y.~Tan, J.~Eledath, and A.~Z. Chaudhry, ``Multiple target
  tracking by integrating track refinement and data association,'' in
  \emph{Proceedings of the 15th International Conference on Information
  Fusion}.\hskip 1em plus 0.5em minus 0.4em\relax IEEE, 2012, pp. 1254--1260.

\bibitem{divakaran2014real}
A.~Divakaran, Q.~Yu, A.~Tamrakar, H.~S. Sawhney, J.~Zhu, O.~Javed, J.~Liu,
  H.~Cheng, and J.~Eledath, ``Real-time object detection, tracking and
  occlusion reasoning,'' May~23 2014, uS Patent App. 14/286,305.

\bibitem{gu2017ava}
C.~Gu, C.~Sun, S.~Vijayanarasimhan, C.~Pantofaru, D.~A. Ross, G.~Toderici,
  Y.~Li, S.~Ricco, R.~Sukthankar, C.~Schmid \emph{et~al.}, ``Ava: A video
  dataset of spatio-temporally localized atomic visual actions,'' \emph{arXiv
  preprint arXiv:1705.08421}, 2017.

\bibitem{he2016deep}
K.~He, X.~Zhang, S.~Ren, and J.~Sun, ``Deep residual learning for image
  recognition,'' in \emph{Proceedings of the IEEE conference on computer vision
  and pattern recognition}, 2016, pp. 770--778.

\bibitem{ren2015faster}
S.~Ren, K.~He, R.~Girshick, and J.~Sun, ``Faster r-cnn: Towards real-time
  object detection with region proposal networks,'' in \emph{Advances in Neural
  Information Processing Systems}, 2015, pp. 91--99.

\bibitem{Wojke2017simple}
N.~Wojke, A.~Bewley, and D.~Paulus, ``Simple online and realtime tracking with
  a deep association metric,'' in \emph{2017 IEEE International Conference on
  Image Processing (ICIP)}, 2017, pp. 3645--3649.

\bibitem{Bewley2016_sort}
A.~Bewley, Z.~Ge, L.~Ott, F.~Ramos, and B.~Upcroft, ``Simple online and
  realtime tracking,'' in \emph{2016 IEEE International Conference on Image
  Processing (ICIP)}, 2016, pp. 3464--3468.

\bibitem{REF08a}
R.-E. Fan, K.-W. Chang, C.-J. Hsieh, X.-R. Wang, and C.-J. Lin, ``Liblinear: A
  library for large linear classification,'' \emph{Journal of Machine Learning
  Research}, vol.~9, pp. 1871--1874, 2008.

\bibitem{zhang2014novel_reid}
Z.~Zhang, Y.~Chen, and V.~Saligrama, ``A novel visual word co-occurrence model
  for person re-identification,'' in \emph{European Conference on Computer
  Vision}.\hskip 1em plus 0.5em minus 0.4em\relax Springer, 2014, pp. 122--133.

\bibitem{Das2014}
A.~Das, A.~Chakraborty, and A.~K. Roy-Chowdhury, ``Consistent re-identification
  in a camera network,'' in \emph{Proceedings of the European Conference on
  Computer Vision}, 2014, pp. 330--345.

\bibitem{Wang2016Person}
T.~Wang, S.~Gong, X.~Zhu, and S.~Wang, ``Person re-identification by
  discriminative selection in video ranking,'' \emph{IEEE Transactions on
  Pattern Analysis and Machine Intelligence}, vol.~38, no.~12, pp. 2501--2514,
  Dec 2016.

\bibitem{Zheng2016Towards}
W.~S. Zheng, S.~Gong, and T.~Xiang, ``Towards open-world person
  re-identification by one-shot group-based verification,'' \emph{IEEE
  Transactions on Pattern Analysis and Machine Intelligence}, vol.~38, no.~3,
  pp. 591--606, March 2016.

\bibitem{Chen2018PAMI_REID}
Y.~C. Chen, X.~Zhu, W.~S. Zheng, and J.~H. Lai, ``Person re-identification by
  camera correlation aware feature augmentation,'' \emph{IEEE Transactions on
  Pattern Analysis and Machine Intelligence}, vol.~40, no.~2, pp. 392--408, Feb
  2018.

\end{thebibliography}
%



%

\begin{IEEEbiography}
[{\includegraphics[width=1in,height=1.25in,clip,keepaspectratio]{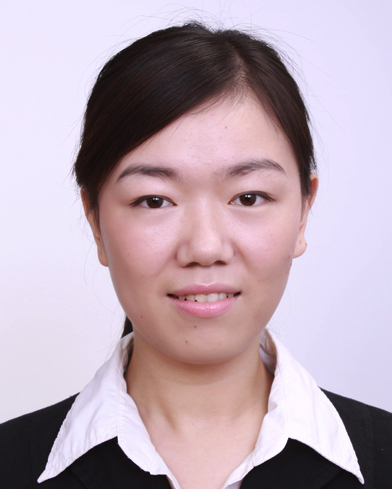}}]{Yuting Chen}
received a B.E. in Automation and M.S. in Signal and Information Processing from Xi'an Jiaotong University, China in 2008 and 2011. She received her Ph.D. in Systems Engineering from Boston University in 2017. Since 2018, she has been a data scientist at Adobe. Her research interests include similarity learning, video retrieval, object detection and recognition.
\end{IEEEbiography}



\begin{IEEEbiography}
[{\includegraphics[width=1in,height=1.25in,clip,keepaspectratio]{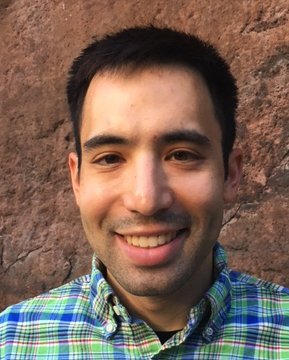}}]
{Joseph Wang}
received the B.S. from Columbia University in 2008 and M.S. and Ph.D. from Boston University in 2014. He was appointed as a Research Professor at Boston University in 2015. Since 2017, he has been a research scientist at Amazon. His main areas of research interest are large-scale learning and test-time efficient learning.
\end{IEEEbiography}

\begin{IEEEbiography}
[{\includegraphics[width=1in,height=1.25in,clip,keepaspectratio]{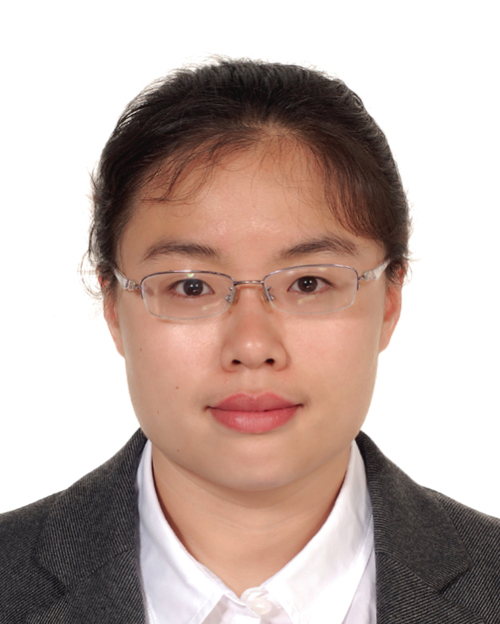}}]
{Yannan Bai}
is a graduate student in Electrical and Computer Engineering at the Collenge of Engineering, Boston University. Her research interests include video processing and computer vision. Bai received her BS in information engineering from Shanghai Jiao Tong University.
\end{IEEEbiography}

\begin{IEEEbiography}
[{\includegraphics[width=1in,height=1.25in,clip,keepaspectratio]{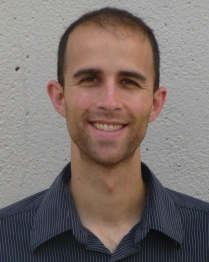}}]
{Gregory~Casta{\~n\'o}n}
is a research scientist at Systems and Technology Research in Woburn, Massachusetts.  He received his PhD in Electrical Engineering from Boston University in 2016 in large scale activity detection.  His interests include multi-target tracking, information fusion, activity recognition and detection.
\end{IEEEbiography}

\begin{IEEEbiography}
[{\includegraphics[width=1in,height=1.25in,clip,keepaspectratio]{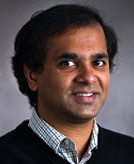}}]
{Venkatesh Saligrama}
is a faculty member in the Department of Electrical and Computer
Engineering and Department of Computer Science (by courtesy) at Boston University.
He holds a PhD from MIT. His research interests are in Statistical Signal Processing, Machine
Learning and Computer Vision, Information and Decision theory. He has edited a book on
Networked Sensing, Information and Control. He has served as an Associate Editor for IEEE
Transactions on Information Theory, IEEE Transactions on Signal Processing and has been on
Technical Program Committees of several IEEE conferences. He is the recipient of numerous
awards including the Presidential Early Career Award (PECASE), ONR Young Investigator
Award, the NSF Career Award and a NIPS 2014 workshop best student paper award on Analysis of Ranking Data. More information about his work is available at \url{http://sites.bu.edu/data}.
\end{IEEEbiography}





\end{document}